\documentclass{aa}

\usepackage[dvipsnames]{xcolor} 
\usepackage{graphicx, subfigure, amsmath, pifont}
\usepackage{amssymb}
\usepackage{multicol}
\usepackage[breaklinks=true]{hyperref} 
\hypersetup{colorlinks=true,citecolor=Blue}
\usepackage{natbib}
\bibpunct{(}{)}{;}{a}{}{,} % to follow the A&A style
\usepackage[english]{babel}
\usepackage{blindtext}
\usepackage[normalem]{ulem}
\usepackage{comment}
\usepackage{lmodern}
\usepackage{ulem} 
\usepackage{soul}

\let\oldpageref\pageref
\renewcommand{\pageref}{\oldpageref*}

\graphicspath{{./}{figures/}}

\begin{document} 

\title{Binary orbit and disks properties of the RW\,Aur system using ALMA observations}

   \author{
    N.~T.~Kurtovic\inst{\ref{mpe}},
    S.~Facchini\inst{\ref{unimi}}, 
    M.~Benisty\inst{\ref{ipag}},
    P.~Pinilla\inst{\ref{uclondon}},
    S.~Cabrit\inst{\ref{ipag},\ref{sorbonne}}, 
    E.~L.~N.~Jensen\inst{\ref{dpasc}}, 
    C.~Dougados\inst{\ref{ipag}}, 
    R.~Booth\inst{\ref{leeds},\ref{imperial}}, 
    C.~N.~Kimmig\inst{\ref{ita}}, 
    C.~F.~Manara\inst{\ref{eso}}, 
    J.~E.~Rodriguez\inst{\ref{msu}}
   }
   \institute{
   Max-Planck-Institut f\"ur Extraterrestrische Physik, Giessenbachstrasse 1, 85748 Garching, Germany. \label{mpe}  \\ \email{kurtovic@mpe.mpg.de}
   \and Dipartimento di Fisica, Universitá degli Studi di Milano, via Celoria 16, Milano, Italy. \label{unimi}
%   \and Unidad Mixta Internacional Franco-Chilena de Astronom\'{i}a (CNRS UMI 3386), Departamento de Astronom\'{i}a, Universidad de Chile, Camino El Observatorio 33, Las Condes, Santiago, Chile \label{cnrs}
   \and Univ. Grenoble Alpes, CNRS, IPAG, F-38000 Grenoble, France. \label{ipag}
   \and Mullard Space Science Laboratory, University College London, Holmbury St Mary, Dorking, Surrey RH5 6NT, UK.\label{uclondon}
   \and Sorbonne Université, Observatoire de Paris, Université PSL, CNRS, LERMA, F-75014 Paris, France. \label{sorbonne}
   \and Department of Physics \& Astronomy, Swarthmore College, Swarthmore PA 19081, USA. \label{dpasc}
   \and School of Physics and Astronomy, University of Leeds, Leeds, LS2 9JT, UK. \label{leeds}
   \and Astrophysics Group, Department of Physics, Imperial College London, Prince Consort Rd, London SW7 2AZ, UK. \label{imperial}
   \and Institute for Theoretical Astrophysics, Zentrum für Astronomie, Heidelberg University, Albert Ueberle Str. 2, 69122 Heidelberg, Germany. \label{ita}
   \and European Southern Observatory, Karl-Schwarzschild-Str. 2, D-85748 Garching bei München, Germany. \label{eso}
   \and Center for Data Intensive and Time Domain Astronomy, Department of Physics and Astronomy, Michigan State University, East Lansing, MI 48824, USA. \label{msu}
   }
   \date{}

% F.M. 0000-0002-1637-7393

 \authorrunning{N.T.~Kurtovic}
 \titlerunning{RW Aur}
% \abstract{}{}{}{}{} 
% 5 {} token are mandatory

  \abstract
  % context heading (optional)
    {The dynamical interactions between young binaries can perturb the material distribution of their circumstellar disks, and modify the planet formation process. In order to understand how planets form in multiple stellar systems, it is necessary to characterize both their binary orbit and their disks properties.} 
  % aims heading (mandatory)
    { In order to constrain the impact and nature of the binary interaction in the RW\,Aur system (bound or unbound), we analyzed the circumstellar material at 1.3\,mm wavelengths, as observed at multiple epochs by the Atacama Large (sub-)millimeter Array (ALMA). }
  %methods heading (mandatory)
    {We analyzed the disk properties through parametric visibility modeling, and we used this information to constrain the dust morphology and the binary orbital period. } 
  % results heading (mandatory)
    {We imaged the dust continuum emission of RW\,Aur with a resolution of 3\,au, and we find that the radius enclosing 90\% of the flux ($R_{90\%}$) is 19\,au and 14\,au for RW\,Aur\,A and B, respectively. By modeling the relative distance of the disks at each epoch, we find a consistent trend of movement for the disk of RW\,Aur\,B moving away from the disk of RW\,Aur\,A at an approximate rate of 3\,mas\,yr$^{-1}$ (about 0.5\,au\,yr$^{-1}$ in sky-projected distance). 
    By combining ALMA astrometry, historical astrometry, and the dynamical masses of each star, we constrain the RW\,Aur binary stars to be most likely in a high-eccentricity elliptical orbit with a clockwise prograde orientation relative to RW\,Aur\,A, although low-eccentricity hyperbolic orbits are not ruled out by the astrometry. 
    Our analysis does not exclude the possibility of a disk collision during the last interaction, which occurred $295_{-74}^{+21}$\,yr ago relative to beginning of 2024. Evidence for the close interaction is found in a tentative warp of 6\,deg in the inner 3\,au of the disk of RW\,Aur\,A, in the brightness temperature of both disks, and in the morphology of the gas emission. A narrow ring that peaks at 6\,au around RW\,Aur\,B is suggestive of captured material from the disk around RW\,Aur\,A.} 
  % conclusions heading (optional), leave it empty if necessary  {}
    %{}
    {}

  \keywords{protoplanetary disks, stars: binaries (close), techniques: high angular resolution}

  \maketitle

%\onecolumn

\section{Introduction}\label{sec:intro}

The dynamical interactions between young stellar systems can significantly impact their planet formation environment. Binaries are known to truncate the disks of their companions \citep[e.g.][]{papaloizou1977, artymowicz1994, manara2019, ragusa2021, rota2022, cuello2023, zagaria2023, zurlo2023}, disrupt the disk material into highly eccentric or unbound orbits \citep[e.g.][]{rodriguez2018}, and modify the material distribution over the disk, generating spirals, arc-like structures, and inducing warps \citep[e.g.][]{kurtovic2018, Nealon2020}. Therefore, it is crucial to study young binaries undergoing these processes for understanding the planet population in multiple stellar systems \citep{offner2023}. 

Close encounters in highly eccentric systems or unbound fly-bys are more common in the early stages of star and disk formation \citep{pfalzner2013, Bate2018}, and these interactions generate structures that only last for a short astronomical time \citep[scales of thousands of years, e.g.,][]{Cuello2019}. However, the consequences of these encounters can be catastrophic for the disk and its potential planetary system. 
A few systems show evidence of interaction with a companion, such as RW\,Aur \citep{Cabrit2006, rodriguez2018}, SR24 \citep{mayama2010, fernandez2017}, HV\,Tau with DO\,Tau \citep{winter2018b}, FU\,Ori \citep{takami2018}, AS\,205 \citep{kurtovic2018}, BHB2007-11 \citep{alves2019}, UX\,Tau \citep{menard2020, Zapata2020}, and Z\,CMa \citep{Dong2022}, with the first system being the focus of this work.

The RW\,Aur system is composed of at least two stars located at 154\,pc away from us \citep{gaia2016b,gaia2021edr3}, each hosting its own disk \citep{Cabrit2006,rodriguez2018,long2019}. However, RW\,Aur\,A has been suspected of being a spectroscopic binary \citep{gahm1999}. 
The luminosities of stars A and B are 0.88$\,L_\odot$ and 0.53$\,L_\odot$, respectively \citep[][corrected to 154\,pc]{Herczeg2014}, even though RW\,Aur\,A is known for having a variable luminosity and dimming events, during which the optical brightness can change by as much as 2\,mag during periods of several months \citep[][]{Chou2013, petrov2015}. These variability events have been hypothesized to be related to dusty inner disk winds \citep{Shenavrin2015, Bozhinova2016, Koutoulaki2019} and tidally disrupted material or disk misalignments \citep{Rodriguez2013, rodriguez2016, Dai2015, facchini2016}. Evidence of a tidal interaction was directly identified by \citet{Cabrit2006} using the IRAM Plateau de Bure Interferometer, with the detection of an arc-like emission in the $^{12}$CO J=2-1 molecular line. A later follow-up with the Atacama Large (sub-)millimeter Array (ALMA) by \citet{rodriguez2018} found multiple additional $^{12}$CO features and suggested that the RW\,Aur system has undergone multiple encounters, an hypothesis that has been discussed in additional works \citep[e.g.,][]{Dai2015,Dodin2019}. 

The impact of the phenomena described above on the circumstellar disk of each binary has remained mostly unclear, because the dust continuum disks were barely resolved by \citet{rodriguez2018}, and the orbital parameters have not yet been constrained from the available optical observations. Thus, the present work aims to study the nature and impact of the binary interaction in the distribution of the gas and dust emission in the individual disks. 
This study uses ALMA observations at 1.3\,mm at high angular resolution, which are further described in Sect.~\ref{sec:observations}. We use these datasets to recover the gas and dust continuum emission morphology, as shown and analyzed in Sect.~\ref{sec:results}. Our findings are discussed in Sect.~\ref{sec:discussion}, and the conclusions are presented in Sect.~\ref{sec:conclusions}.

%%%%%%%%%%%%%%%%%%%%%%%%%%%%%%%%%%%%%%%%%%%%%%%%%%%%%%%%%%%%%%%%%%%%%%%%%%%%%%%%

%% OBSERVATIONAL DATA DESCRIPTION
\section{Observations} \label{sec:observations}

\begin{table*}[t]
\centering
\caption{ALMA observations of RW\,Aur. }
\begin{tabular}{ c|c|c|c|c|c|c|c } 
  \hline
  \hline
\noalign{\smallskip}
Project & Code & PI Name & Obs Date & N antennas & Baselines & Exp Time & Freq \\
code    &      &         &          &            & (m)       & (min)    & (GHz) \\
\noalign{\smallskip}
  \hline
\noalign{\smallskip}
2015.1.01506.S & SB1 & Rodriguez, J. & 2016-09-29 & 39 & 15 - 3248  & 32.38 & 217.5 - 232.6 \\
               &     &                   & 2016-09-29 & 39 & 15 - 3248  & 32.38 & \\
               &     &            	      & 2016-09-30 & 39 & 15 - 3144  & 32.38 & \\
               &     &       	     	 & 2016-09-30 & 39 & 15 - 3144  & 32.38 & \\
\noalign{\smallskip}
  \hline
\noalign{\smallskip}
2016.1.00877.S & SB2 & Rodriguez, J. & 2016-12-08 & 11 & 9 - 45     & 25.54 & 217.5 - 232.6 \\
               &     &       	         & 2016-12-08 & 11 & 9 - 45     & 25.54 & \\
               &     &              	 & 2016-12-08 & 11 & 9 - 45     & 25.54 & \\
\noalign{\smallskip}
  \hline
\noalign{\smallskip}
2016.1.01164.S & SB3 & Herczeg, G.  & 2017-08-31 & 45 & 21 - 3697  & 8.97  & 217.0 - 234.0 \\
\noalign{\smallskip}
  \hline
\noalign{\smallskip}
2017.1.01631.S & SB4 & Facchini, S. & 2018-12-02 & 46 & 15 - 952   & 32.92 & 215.6 - 233.4 \\
        	   &     &                   & 2018-12-05 & 47 & 15 - 784   & 32.92 & \\
\noalign{\smallskip}
  \hline
\noalign{\smallskip}
2018.1.00973.S & LB1 & Facchini, S. & 2017-10-08 & 49 & 41 - 16196 & 38.73 & 217.6 - 232.9 \\
	           &     &                   & 2017-10-09 & 51 & 41 - 16196 & 38.66 & \\
	           &     &            	     & 2017-10-18 & 51 & 41 - 16196 & 39.82 & \\
\noalign{\smallskip}
  \hline
  \hline
\end{tabular}
\label{tab:obs_log}  
\end{table*}

% Data description
This work includes 1.3\,mm observations of the system RW\,Aur from several different ALMA projects, that are listed in Table~\ref{tab:obs_log}, with an approximate time-span of 2 years and 2 months. We name each project with an identification code for the extension of the antenna baselines: SB for the observations performed with compact antenna configurations (short baselines) and LB for extended configurations (long baselines). The projects SB1 and SB2 were already published in \citet{rodriguez2018}, while SB3 was part of the Taurus survey published in \citet{long2019} and \citet{manara2019}. For the datasets SB4 and LB1, which have not been published before, the correlator was configured to observe five and four spectral windows, respectively. SB4 contains two spectral windows covering dust continuum emission centered at $218.503$\,GHz and $232.003$\,GHz, and the remaining three spectral windows were centered at the molecular lines $^{12}$CO, $^{13}$CO, and C$^{18}$O in their transition J=2-1. The frequency resolution of the continuum is $1128.91$\,kHz, and for $^{12}$CO, it is 141.11\,kHz, and it is 282.23kHz for the remaining two lines. The LB1 observation, on the other hand, has three spectral windows for the continuum and one spectral window for $^{12}$CO (J=2-1). The frequency resolution of all of them is $1128.91$\,kHz, which is about 1.3\,km\,s$^{-1}$ at $230.538$\,GHz.

% Continuum self calibration
We started from the pipeline-calibrated measurement set generated with the \texttt{scriptforPI} delivered by ALMA. Using \texttt{CASA 5.6.2}, we extracted the dust continuum emission from the spectral windows targeting gas emission lines, and to do this, we flagged the channels located at $\pm 25\,$km\,s$^{-1}$ from the system approximate velocity at the local standard of rest (VLSR), which is about 6\,km\,s$^{-1}$ for RW\,Aur. 
The remaining channels were combined with the other continuum spectral windows to obtain a pseudo-continuum dataset, and we averaged into 125\,MHz channels and $6\,$s bins to reduce data volume. The signal-to-noise ratio (S/N) for each project was sufficient to self-calibrate each by itself, and therefore, we did not align the different observations to the same phase center prior to self-calibration. 
Each project was completely observed within days between the first and last observation. We therefore treated each observation as a single epoch. The imaging for the self-calibration process was made with a Briggs robust parameter of 0.5, except for LB1, where we used 1.0. 
This higher value was chosen to increase the S/N of the CLEAN image, to allow us to recover more flux in the CLEAN model. 
We combined all the scans and spectral windows for each \texttt{gaincal} execution. The reference dust continuum image at high angular resolution, that we produced from the observation LB1, has an angular resolution of $18\times30$\,mas. 

\begin{figure*}[t]
 \centering
        \includegraphics[width=18cm]{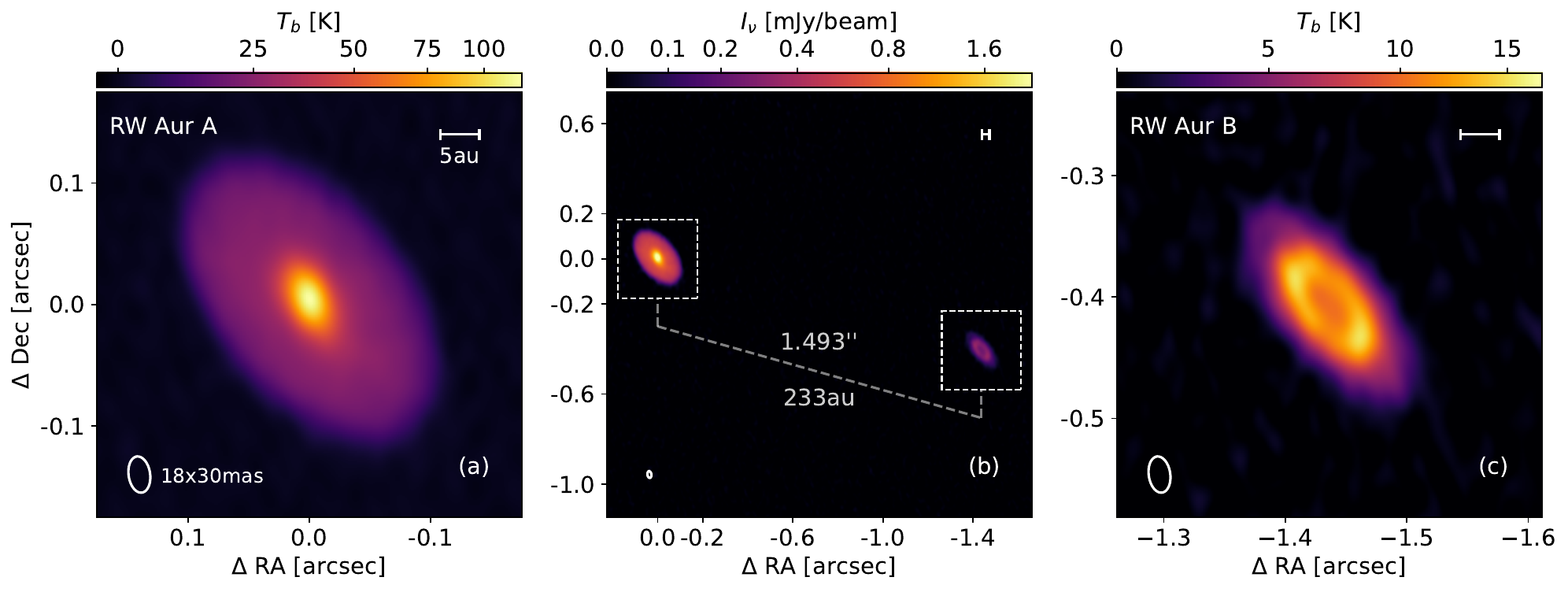}\\
        \vspace{-0.1cm}
   \caption{Dust continuum emission imaged with observation LB1. Panels (a) and (c) show each disk individually in brightness temperature. The panel (b) shows both disks together in brightness per beam, and the projected distance at this epoch. The beam resolution is the same for all the panels, and its size is $18\times30$\,mas, as shown in the lower left corner of panel (a). The scale bar is 5\,au at the distance of the source.}
   \label{fig:gallery_cont}
\end{figure*}

% Gas calibration
The calibration tables obtained from the dust continuum self-calibration of each observation were then applied to the original measurement sets, which contained dust continuum and molecular line emission. We subtracted the continuum emission with the task \texttt{uvcontsub} and obtained a measurement set for each molecular line for each epoch. The $^{12}$CO line is present in all the observations except for SB3, while $^{13}$CO and C$^{18}$O are present in SB1, SB2, SB3 and SB4. We combined the visibilities of different epochs to generate the gas images of each tracer under the assumption that spatial movement is negligible at the scales covered by the angular resolution of the gas observations (about 6\,mas over the two-year period; see Sect.~\ref{sec:results_uvmodel}). 
The $^{12}$CO line is the brightest gas emission line available in our data and was imaged twice. The first image combined the observations SB1, SB2, and SB4 to optimize the sensitivity to large spatial scales (which we call SB $^{12}$CO image), and the second image combined all observations to maximize the sensitivity at high angular resolution (which we call LB $^{12}$CO image). The $^{12}$CO images do not include the SB3 observation, which did not targeted this line. 
When combined with the LB1 dataset, the highest allowed velocity resolution is 1.3\,km\,s$^{-1}$, while the SB $^{12}$CO image was imaged with 0.5\,km\,s$^{-1}$ to optimize the balance between sensitivity and velocity resolution. The $^{13}$CO and C$^{18}$O images have the same imaging setup as the SB $^{12}$CO image, but they include SB3. The angular resolution of the SB $^{12}$CO image is $267\times414$\,mas, and the resolution in the LB $^{12}$CO image is $69\times98$\,mas. 

% Imaging + JvM
The gas and continuum emission were imaged using the task \texttt{tclean}. To avoid introducing Point-Spread-Function (PSF) artifacts that could be mistaken for faint emission, we lowered the \texttt{gain} parameter to $0.05$ and increased the \texttt{cyclefactor} to $1.5$, for more conservative imaging compared to the default values\footnote{Check \url{https://casadocs.readthedocs.io/en/stable/api/tt/casatasks.imaging.tclean.html\#description} for a description of the parameters}, and we cleaned down to the 4$\sigma$ threshold. 
We applied the JvM correction to our images, which scales the residuals to account for the volume ratio $\epsilon$ between the PSF of the images and the restored Gaussian of the {\small{\texttt{CLEAN}}} beam, as described in \citet{jorsater1995} and \citet{czekala2021}. 
We used the package \texttt{bettermoments} \citep{bettermoments, teague2019} to create additional image products from the channel maps. We calculated the peak intensity image by fitting a quadratic function in each pixel along the velocity axis, which also yielded the velocity associated with the peak flux. The same package was also used to generate the moment 0 and moment 1 of each velocity cube. All the moment images were clipped at $0\sigma$ (negative emission is removed), and no mask was used. An additional clipped image was generated from the LB $^{12}$CO image, where we only considered pixels with emission over 1.3\,mJy\,beam$^{-1}$\,km\,s$^{-1}$, with the aim of filtering extended gas emission and recovering the bright localized emission from the disks\footnote{All the selfcalibrated products, including measurement sets and fits files (with and without JvM correction) are available for download at \url{https://doi.org/10.5281/zenodo.12825068}.}. %This value was chosen as the $^{12}$CO plateaus 

% data treatment for uvmodeling
We applied visibility modeling to the continuum visibilities of each epoch. To further reduce the data volume after completing the self-calibration, we averaged the continuum emission into one channel per spectral window and 24\,s. We used the central frequency of each binned channel to convert the visibility coordinates into wavelength units, and we did not combine the visibility tables of different epochs.

%%%%%%%%%%%%%%%%%%%%%%%%%%%%%%%%%%%%%%%%%%%%%%%%%%%%%%%%%%%%%%%%%%%%%%%

\section{Results}
\label{sec:results}

\subsection{Dust continuum emission}

The dust continuum observations are resolved into two independent disks \citep[as it had been observed before in][]{Cabrit2006, rodriguez2018, long2019, manara2019}, one disk around each source of the system, as shown in Fig.~\ref{fig:gallery_cont}. 
When imaged at very high angular resolution, RW\,Aur\,A is resolved into a compact ($R_{90\%}=0.124''$ or 19\,au, with $R_{90\%}$ the radius that encloses 90\% of the dust continuum emission) centrally peaked disk, without evidence of an annular ring-like structure with the $3\times5\,$au beam resolution. Located about 233\,au in projected distance to the southwest is RW\,Aur\,B, which is also a compact disk ($R_{90\%}=0.093''$ or 14\,au) with evidence of a dust continuum ring. The sizes were constrained through visibility modeling of the dust continuum emission visibilities, as described in Sect.~\ref{sec:results_uvmodel}.

\begin{figure}[t]
 \centering
        \includegraphics[width=7.5cm]{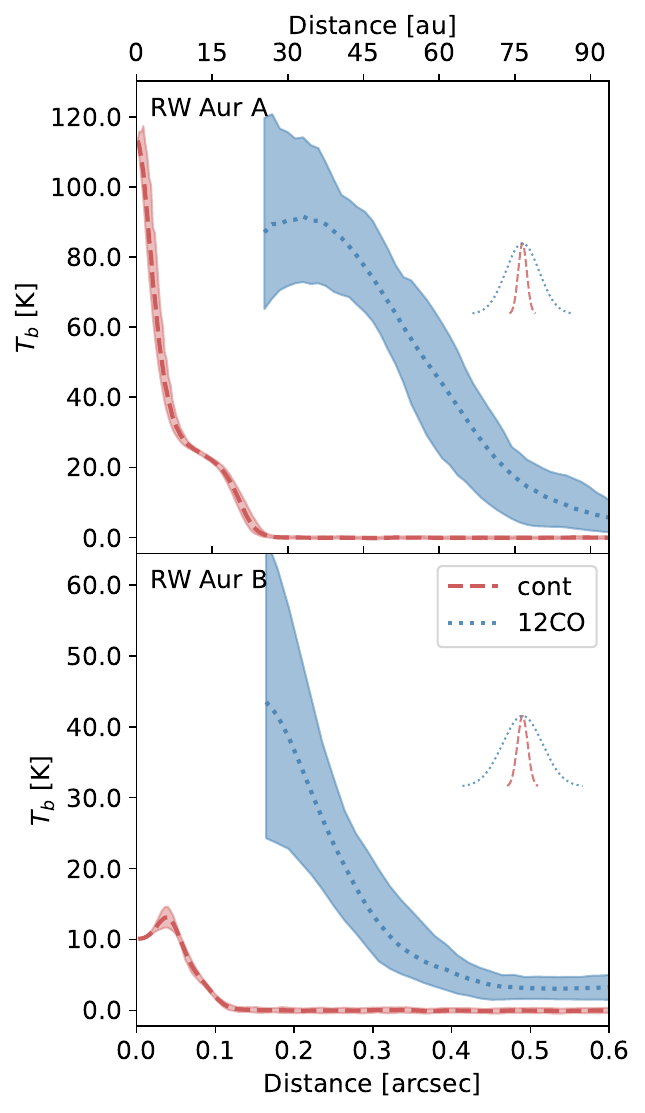}\\
        \vspace{-0.1cm}
   \caption{Azimuthally averaged brightness temperature profile calculated from the CLEAN image of the dust continuum shown as dashed red lines. The brightness temperature profile for the $^{12}$CO is shown as a dotted blue line, as measured from the peak brightness LB image. The inner three beams have been excluded due to the brightness dilution in the channel maps. 
   The colored regions show the 1$\sigma$ dispersion at each location. 
   The Gaussians in the right part of each panel represent the average radial resolution of the dust continuum image and LB $^{12}$CO. The beam sizes are shown in Fig.~\ref{fig:gallery_cont} and \ref{fig:gallery_co}}. 
   \label{fig:profiles}
\end{figure}

We constrained the flux of each source from the visibility modeling by integrating it from the model images. The dust continuum of RW\,Aur\,A is almost eight times brighter than that of its companion RW\,Aur\,B, with $34.5\,$mJy and $4.4$\,mJy of integrated flux, respectively, as also shown in Table~\ref{tab:disk_prop}. RW\,Aur\,A is also considerably hotter in brightness temperature, with a peak $T_b=120$\,K in the disk center. The brightness temperature profile of RW\,Aur\,A decreases monotonically as a function of radii and remains higher than 20\,K until a radius of $107\,$mas (or 17\,au). In contrast, the maximum brightness temperature of RW\,Aur\,B at any radii is 15\,K. The azimuthally averaged brightness temperature profiles are shown in Fig.~\ref{fig:profiles}, and their amplitude difference is further discussed in Sect.~\ref{sec:discussion:cont_structure}.

For an estimate of the dust mass, we followed \citet{hildebrand1983}:
\begin{equation}
    M_{\text{dust}} = \frac{d^2\,F_\nu}{\kappa_\nu\,B_\nu(T(r))} \text{,}
\label{eq:mdust}
\end{equation}
\noindent where $d$ is the distance to the source, $\nu$ is the observed frequency, $B_\nu$ is the Planck function at the frequency $\nu$, and $\kappa_\nu=2.3(\nu/230\,\text{GHz})^{0.4}\,\text{cm}^{2}\text{g}^{-1}$ is the frequency-dependent mass absorption coefficient \citep[as in][]{andrews2013}. 
The additional assumption for this calculation was that the dust emission at 1.3\,mm is emitted by optically thin dust with a known temperature, that is commonly set to 20\,K for standard reference \citep[as in ][]{ansdell2016, cieza2019}. As the brightness temperature is higher than 20\,K for about 75\% of the emitting area of RW\,Aur\,A, the assumption of optically thin emission fails, and we can therefore only provide a lower limit to its dust mass. 
When we assume a midplane temperature of 20\,K (for comparison with other surveys and observations), we obtain a dust mass of $23.9\,M_\oplus$ and $3.1\,M_\oplus$ for A and B, respectively. 
Instead, with an average brightness temperature of RW\,Aur\,A within its $R_{90\%}$ of 27\,K, the dust mass becomes $16.4\,M_\oplus$.

\subsection{Continuum visibility modeling} \label{sec:results_uvmodel}

\begin{table}[t]
\centering
\caption{Disks properties and relative astrometry from visibility modeling.}
\begin{tabular}{ c|c|c|c } 
  \hline
  \hline
\noalign{\smallskip}
           & Property      & Best value $\pm 1\sigma$ & unit \\
\noalign{\smallskip}
  \hline
\noalign{\smallskip}
RW\,Aur\,B & $r_{AB}(SB1)$      & $1491.49 \pm 0.15  $ & mas \\
position   & $r_{AB}(SB3)$      & $1493.49 \pm 0.64  $ & mas \\
relative to& $r_{AB}(SB4)$      & $1499.50 \pm 0.60  $ & mas \\
RW\,Aur\,A & $r_{AB}(LB1)$      & $1493.56 \pm 0.15  $ & mas \\
           & $\theta_{AB}(SB1)$ & $254.155 \pm 0.008 $ & deg \\
           & $\theta_{AB}(SB3)$ & $254.047 \pm 0.026 $ & deg \\
           & $\theta_{AB}(SB4)$ & $254.041 \pm 0.033 $ & deg \\
           & $\theta_{AB}(LB1)$ & $254.055 \pm 0.006 $ & deg \\
\noalign{\smallskip}
  \hline
\noalign{\smallskip}
Disks      & inc$_A$ & $54.93 \pm 0.04  $ & deg \\
Geometry   & PA$_A$  & $39.35 \pm 0.05  $ & deg \\
           & inc$_B$ & $63.98 \pm 0.16  $ & deg \\
           & PA$_B$  & $39.65 \pm 0.24  $ & deg \\
\noalign{\smallskip}
  \hline
\noalign{\smallskip}
Disks      & $R_{B,ring}$ & $41.49  \pm 0.29$  & mas \\
Continuum  & $R_{A,68\%}$ & $101.37 \pm 0.13$  & mas \\
Properties & $R_{B,68\%}$ & $ 68.55 \pm 0.23$  & mas \\
           & $R_{A,90\%}$ & $123.77 \pm 0.12$  & mas \\
           & $R_{B,90\%}$ & $ 92.95 \pm 0.14$  & mas \\
           & $F_{A,mm}$   & $34.504 \pm 0.010$ & mJy \\
           & $F_{B,mm}$   & $ 4.427 \pm 0.006$ & mJy \\
\noalign{\smallskip}
  \hline
  \hline
\end{tabular}
\tablefoot{Relative position of RW\,Aur\,B respect to A and disks geometries are free parameters in the MCMC, while the disks continuum properties are products of the fit. The remaining free parameters of our model are in Table~\ref{tab:uv_mcmc}.}
\label{tab:disk_prop}  
\end{table}

To determine the dust continuum emission morphology and properties, we used the packages \texttt{galario} \citep{Tazzari2017} and \texttt{emcee} \citep{emcee2013} to fit a parametric visibility model to the data. We generated one independent image for each source and calculated the visibilities of each model separately. The additive properties of the Fourier transform allowed us to add the visibilities of each source and compare the combination to the observations. 

We fit all the epochs at the same time with the same intensity models. However, we allowed the disk centers to be different in each observation, with the underlying assumption that each disk brightness distribution is constant during the two years covered by our data, and the only possible difference between epochs are the relative disks positions. 

Additionally, it is known that the ALMA flux calibration uncertainties can reach up to 10\%, and even higher in particular cases \citep[as in some of the DSHARP sources,][]{Andrews2018}. We therefore added a free parameter for each epoch, a scalar $i_{obs}\approx1$, which multiplied the whole intensity model, and scaled the possible flux density difference. As a reference, we used observation SB4, which has a fixed $i_{SB4}=1$, and thus, all the remaining $i_{obs}$ scaled the models to match the SB4 flux density. This epoch was chosen because it has the highest sensitivity in short baselines and is neither the brightest nor dimmest observation, as confirmed by the scaling factors $i_{obs}$ in Table~\ref{tab:uv_mcmc}.

We fit azimuthally symmetric models to RW\,Aur\,A, and RW\,Aur\,B, taking the morphology of the \texttt{CLEAN} image and model as a guideline. For RW\,Aur\,A, the function used to describe its brightness profile is the following:
\begin{equation}
    I_A(r) \,=\, f_{A0} \,+\, G(r, f_{A1}, \sigma_{A1}) \,+\, TP(r, f_{A2}, r_{A2}, \alpha, \beta) \,\text{,}
\end{equation}

\noindent where $f_{A0}$ is the flux density of a point source at the disk center, $G$ is a centrally peaked Gaussian with a peak flux of $f_{A1}$ and a standard deviation of $\sigma_{A1}$, and $TP$ is a power law tapered with an exponential decay, described in Sect.~\ref{sec:annex_uv-model}. 
The point and Gaussian components are needed to describe the inner emission of the disk, and the tapered power law was used to describe the monotonically decreasing brightness decay with the possibility of a sharp outer edge, if needed. 

\begin{figure*}[t]
 \centering
        \includegraphics[width=18cm]{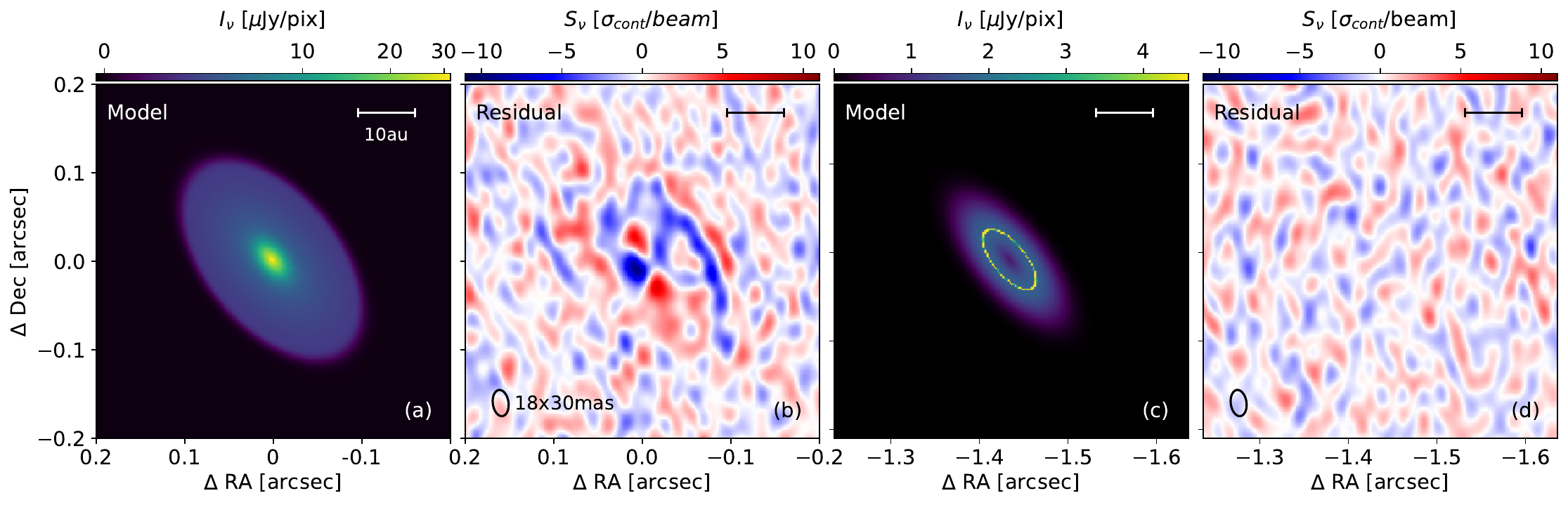}\\
        \vspace{-0.1cm}
   \caption{\textbf{Visibility modeling results. Panels (a)} and \textbf{(c)} show the best parametric models for RW\,Aur\,A and B, respectively. \textbf{Panels (b)} and \textbf{(d)} show the image of the residual visibilities in units of the image sensitivity. The scale bar is 10\,au at the distance of the source, and the beam size in the residual images is shown in the lower left corner.}
   \label{fig:uvmodel_cont}
\end{figure*}

For RW\,Aur\,B, the disk is highly inclined and compact, and therefore, the information of its cavity is limited. A possible parameterization for the dust continuum ring could have been made with a broken Gaussian (i.e. a Gaussian ring with different widths for each side of its peak). 
However, this model returns a very steep inner edge to try to compensate for the slow decrease of its outer edge \citep[similar to the inclined disk MHO\,6 in][]{kurtovic2021}. This problem with the broken Gaussian description can be overcome by slightly increasing the complexity of the model to two Gaussian rings. These Gaussians can become a centrally peaked emission while allowing the model to fit radially asymmetric rings. The equation that describes the model of RW\,Aur\,B as a function of radii is:
\begin{equation}
    I_B(r) \,=\, G(r - r_{B1}, f_{B1}, \sigma_{B1}) \,+\, G(r - r_{B2}, f_{B2}, \sigma_{B2}) \,\text{,}
\end{equation}

\noindent where $f_{B1}$ and $f_{B2}$ are the peak intensity of each ring, centered at $r_{B1}$ and $r_{B2}$ with a Gaussian width of $\sigma_{B1}$ and $\sigma_{B2}$, respectively.

We ran a Markov Chain Monte Carlo (MCMC) fitting on all five epochs at the same time, to achieve complete visibility coverage starting from the 7m antennas of the ACA array from SB2, to the longest baselines from ALMA antenna configuration C-10 with LB1. 
We used a flat prior over the allowed parameter space, and the boundaries for each free parameter were wide enough such that walkers never interacted with them. 
The pixel size for the model images was initially 4\,mas, and we also tested the stability of the fit by running the same models with a pixel size of 2\,mas, from which we obtained consistent results. A summary of the main results is given in Table~\ref{tab:disk_prop}, and the remaining free parameters of the MCMC are shown in Table~\ref{tab:uv_mcmc}.

By construction, the emission from RW\,Aur\,A is centrally peaked, and our model described the disk emission as a monotonically decreasing profile. When the residual visibilities were imaged (see Fig.~\ref{fig:uvmodel_cont}), we found our flat-disk model described most of the structure detected in observation LB1, because the highest peak residual is only $6\sigma$ (compared to $>300\sigma$ of the dust continuum image). 
Even though the residuals contrast are low, their structure suggests that our description of a Gaussian with a point source for the central emission is incomplete. A more detailed discussion of the residuals is given in Sect.~\ref{sec:warp_fit}. 

Our model for RW\,Aur\,B, on the other hand, completely describes the disk emission to the noise level, and no structured residual is detected at the position of the source (see Fig.~\ref{fig:uvmodel_cont}). We find that the RW\,Aur\,B ring peaks at about 42\,mas, which is 6.5\,au from the disk center. Interestingly, both disks show a similar geometry in our line of sight, and their position angles are the same to the uncertainty level. 
RW\,Aur\,B is slightly more inclined than A, as was also estimated by \citet{rodriguez2018} and \citet{manara2019}. Under the assumption that the angular momentum vector of both disks points to the same side of the sky plane, we find a misalignment of $9.15$\,deg, or $118.8$\,deg if they point in different directions.

\subsection{RW Aur B orbit: With ALMA astrometry} \label{sec:rwaurb_orbit}

\begin{figure*}[t]
 \centering
        \includegraphics[width=1.0\textwidth]{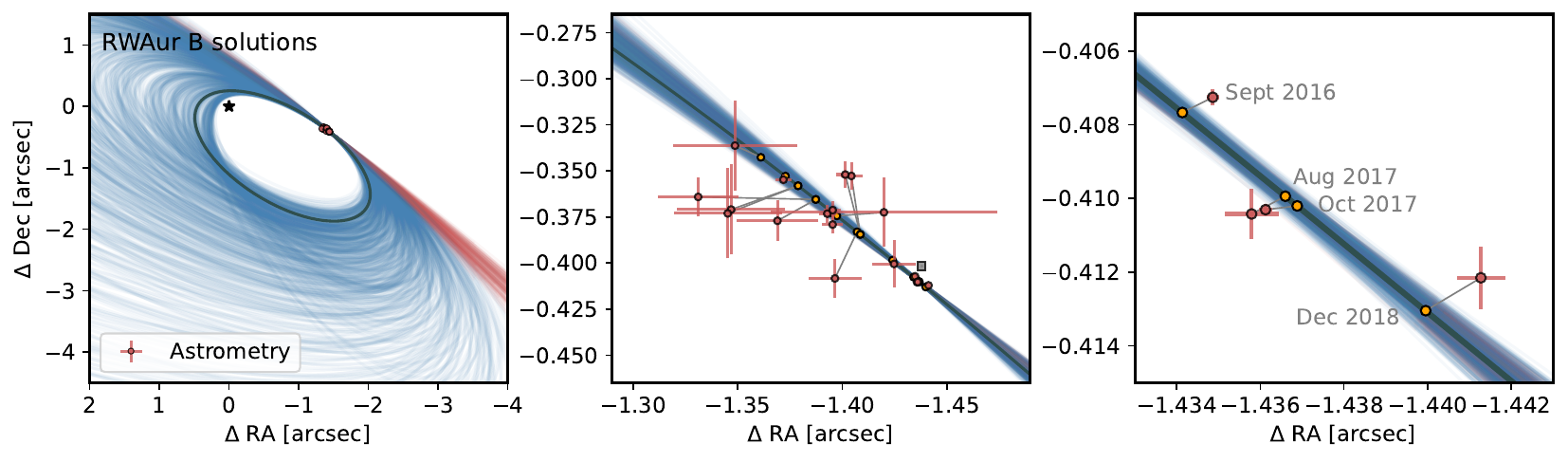}\\
        \vspace{-0.1cm}
   \caption{ 1000 randomly sampled orbital solutions are shown from a larger field of view in the left panel, to a small zoomed region in the right panel. The solid dark line shows the solution with highest likelihood. The blue orbits are elliptical solutions, and the red orbits are hyperbolic solutions. 
   The astrometry considered in this work is shown with red markers, and the position of RW\,Aur\,B with best model is shown as orange circles at the same epoch as the astrometric measurements. 
   For reference, \textit{Gaia} DR3 astrometry is shown with a gray square in the middle panel. The uncertainty from \textit{Gaia} is smaller than the symbol size. }
   \label{fig:rwaurb_pos}
\end{figure*}

\begin{figure*}[t]
 \centering
        \includegraphics[width=1.0\textwidth]{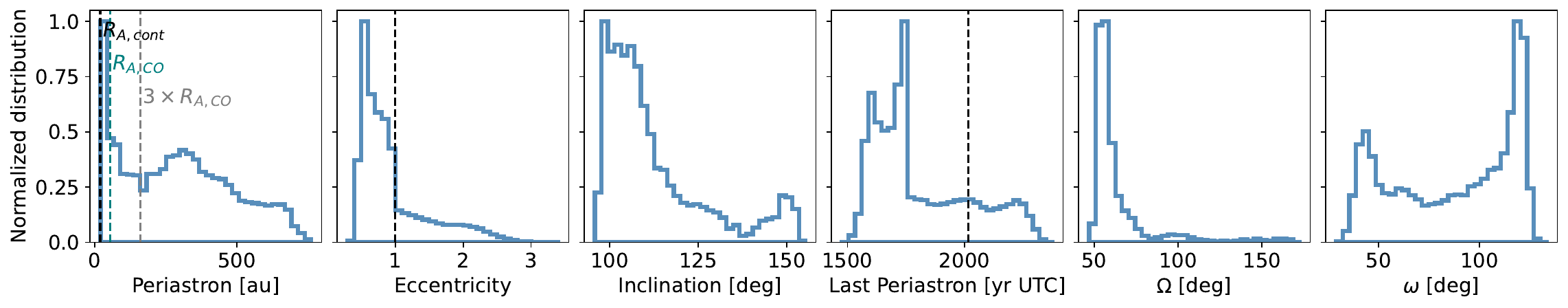}\\
        \vspace{-0.1cm}
   \caption{Normalized probability density distribution for the fitted orbital parameters of RW\,Aur\,B around RW\,Aur\,A. 
   For reference, the dashed lines in the left panel show the continuum, gas, and three times the gas size of RW\,Aur\,A, respectively. The last periastron year is shown in UTC, and the dashed line shows the time of the last observation. }
   \label{fig:orbit_prob_dens}
\end{figure*}

RW\,Aur\,B shows a consistent trend of movement toward the southwest as a function of time. Assuming that the centers of the disks coincide with the location of the stars, we can use the visibility-modeled disk center as a precise astrometric measurement for each system. The uncertainty is considered to be the region including 68\% of the MCMC solutions. This assumption implies that the disks are axisymmetric and have zero eccentricity, and we further discuss this in Sect.~\ref{sec:discussion:astrometry_alma}.

We combined our ALMA astrometry with the historical separation between RW Aur A and B as compiled by \citet{csepany2017}, covering epochs from 1944 to 2013. We excluded two outlier measurements from the historical astrometric positions: an observation from 1991 published in \citet{Leinert1993} that considerably deviates from the positional trend, and is inconsistent by about $0.1''$ in separation and more than $2$ deg in position angle with measurements from 1990 and 1994; and a measurement from 1944 by \citet{joy1944} that does not provide uncertainties. 

We included a single radial velocity (RV) measurement using the relative line-of-sight velocity obtained with ALMA, assigning an average epoch considering our four ALMA observations with equal weighting. 
This radial velocity was assigned to RW\,Aur\,B with a value of $-0.954$ km s$^{-1}$ (where negative means approaching us), relative to the reference rest frame of RW\,Aur\,A. 
Due to the high S/N of the emission of each disk in the $^{12}$CO channel maps, the integrated velocity images, such as moment 1 and velocity at peak brightness, show a ``channelization problem'', where the low-frequency resolution produces a discrete velocity distribution instead of a continuous rotation map (the $^{12}$CO emission is further discussed in Sect.~\ref{sec:obs_gas}). This effect makes it challenging to obtain a meaningful uncertainty from the ALMA cube, and we therefore set a conservative $1\sigma$ dispersion to 100\,m\,s$^{-1}$ for the RV fitting.

To recover the orbit of RW Aur B around RW Aur A, we considered two different eccentricity regimes: elliptical orbits ($\text{ecc}<1$), and hyperbolic orbits ($\text{ecc}>1$). Both families of orbits were calculated using the Python package \texttt{rebound} \citep{rein2012, rein2015}. 
We created a simulation setup with a central mass at the origin of the coordinate system containing the cumulative mass of the binaries, and mass-less particles representing the position of RW Aur B. This is equivalent to a system in which each star has its own mass, and both move relative to an inertial coordinate system. The package \texttt{rebound} receives the true anomaly as input to set the position of a particle relative to the given orbital parameters, instead of physical time. Further details of the time-to-anomaly transformation for elliptical and hyperbolic scenarios are reported in Sect.~\ref{app:sec:hyperbolic}.

We sampled the probability density distribution of the orbital parameters using MCMC with \texttt{emcee} \citep{emcee2013}. The orbits were allowed to be either elliptical or hyperbolic with a uniform prior for any positive eccentricity, except for a small forbidden region around $|1-\text{ecc}|<10^{-4}$ to avoid parabolic solutions. We fit the distance at periastron ($d_{peri}$) instead of the semi-major (semi-transverse) axis ``$a$'' for elliptical (hyperbolic) orbits. 
By definition, $a=d_{peri}/(1-\text{ecc})$, which diverges for values of $\text{ecc}\approx1$ in either the elliptical or hyperbolic regime, introducing an undesired prior over the allowed eccentricity values. 
We used 24 times as many walkers as the number of free parameters, which were eight for a given orbit: the distance at periastron ($d_{peri}$), the eccentricity ($\text{ecc}$), the inclination ($\text{inc}$), the longitude of the ascending node ($\Omega$), the argument of periastron ($\omega$), the time of the last periastron ($t_{peri}$), the parallax ($\text{plx}$), and the total mass ($\text{mtot}$).

After running for over $10^7$ steps per walker, we did not achieve convergence for each walker separately, in particular, in the low likelihood regime for $\Omega$. The main reason for the delayed convergence is the highly correlated relations between the orbital parameters (as shown in Fig.~\ref{app:fig:corner_all}, \ref{app:fig:corner_ell}, and \ref{app:fig:corner_hyp}). However, the probability density distribution derived from the walkers positions reached stable values after $>10^7$ steps. After this stability was achieved, we ran an MCMC recording only one of every 80 steps to reduce data volume, and we derived the solutions from the last $10^5$ recorded positions for every walker.

The orbital solutions are shown in Fig.~\ref{fig:rwaurb_pos}, and their probability density distribution is presented in Fig.~\ref{fig:orbit_prob_dens}. In the appendix, we include the probability density distribution for each parameter after separating elliptical and hyperbolic solutions (see Fig.~\ref{app:fig:orbit_prob_dens}). 
In Table~\ref{tab:orbital_param}, we detail the best values and their $1\sigma$ uncertainty. 
The orbital solutions with the highest likelihood are found in the eccentricity range between 0.5 and 0.95, but there is a tail of solutions with eccentricities $>1$ that are still allowed with the current astrometry, and thus, the hyperbolic solution is not ruled out.

Elliptical and hyperbolic solutions find different distributions for $d_{peri}$ and $t_{peri}$, as shown separately in Fig.~\ref{app:fig:orbit_prob_dens}. While elliptical solutions show closer $d_{peri}$ over a wider range of $t_{peri}$, most hyperbolic solutions are only consistent with very recent $t_{peri}$ (some even in the near future) and large $d_{peri}$, suggesting larger distances for the fly-by interaction. 
The values for $d_{peri}$ in the hyperbolic regime are considerably larger than three times the gas size of the disks (see also Fig.~\ref{app:fig:orbit_prob_dens}), which makes these orbits less consistent with the truncation scenario, as we discuss further in Sect.~\ref{sec:discussion:orbit}. Additionally, hyperbolic orbits are only allowed for very narrow ranges for $inc$, $\Omega$ and $\omega$.

The $t_{peri}$ located in the near future are not consistent with the brightest tidal arc in the $^{12}$CO emission, which must have resulted from close interaction \citep{Dai2015}. Thus, in Table~\ref{tab:orbital_param}, we also include the uncertainties after filtering the orbital solutions by $t_{peri}$ before the epoch of ALMA observations. 
The distributions after the filtering are also shown in Fig.~\ref{app:fig:orbit_prob_dens}. 
For elliptical orbits, we estimate that the period is $2773_{-366}^{+27100}$\,yr, where the uncertainties represent the $1\sigma$ dispersion from the best solution. The large uncertainty for long-period orbits come from the orbits with $\text{ecc}\approx1$, as the semi-major axis of these orbits diverges.

\begin{table}[t]
\centering
\caption{Orbital parameters of the RW\,Aur binary.}
\begin{tabular}{ c c c c } 
  \hline
\noalign{\smallskip}
 Parameter & Best $\pm1\sigma$ & $t_{peri}>0$ & units \\
\noalign{\smallskip}
  \hline
  \hline
\noalign{\smallskip}
$d_{peri}$   &  $55.0_{-27.0}^{+477.5}$  & $55.0_{-27.0}^{+334.0}$ & au \\
\noalign{\smallskip}
ecc          & $0.787_{-0.284}^{+1.155}$ & $0.787_{-0.283}^{+0.305}$ & - \\
\noalign{\smallskip}
inc          & $129.8_{-29.9}^{+21.2}$   & $129.8_{-26.2}^{+21.3}$ & deg \\
\noalign{\smallskip}
$\Omega$     & $73.8_{-21.0}^{+73.7}$    & $73.8_{-19.1}^{+73.6}$ & deg \\
\noalign{\smallskip}
$\omega$     & $42.3_{-5.4}^{+77.6}$     & $42.3_{-5.4}^{+70.1}$ & deg \\
\noalign{\smallskip}
$t_{peri}$   & $295.4_{-456.7}^{+136.0}$ & $295.4_{-227.0}^{+136.0}$ & yrs \\
\noalign{\smallskip}
  \hline
\noalign{\smallskip}
parallax     & $6.406 \pm 0.044$ & $6.406 \pm 0.044$ & mas \\
\noalign{\smallskip}
$m_{tot}$    & $2.233 \pm 0.05\cdot\sqrt{2}$ & $2.233 \pm 0.05\cdot\sqrt{2}$ & $M_\odot$ \\
\noalign{\smallskip}
  \hline
  \hline
\noalign{\smallskip}
\end{tabular}
\tablefoot{ The values shown are the highest likelihood solutions, and $1\sigma$ deviation for the orbit of RW\,Aur\,B around RW\,Aur\,A. $t_{peri}$ is relative to the last observation of ALMA, with positive values in the past, and negative values for future. Second column shows the $1\sigma$ uncertainty when considering orbits where the periastron is in the past. Parallax and total mass are sampled from Normal distributions with a standard deviation equal to the reported $1\sigma$. Posterior distributions are shown in Fig.~\ref{app:fig:corner_all}. }
\label{tab:orbital_param}
\end{table}

\subsection{RW Aur B orbit: Excluding ALMA astrometry}

Additional tests were run excluding the ALMA astrometry from the orbital fit. We fit the orbital parameters for three different scenarios: (i) Only with the historical astrometry from \citet{csepany2017}, (ii) with historical astrometry and the ALMA radial velocity, and (iii) with historical astrometry, the ALMA radial velocity, and the \textit{Gaia} DR3 astrometry. 
We find that the historical astrometry by itself or the historical astrometry combined with the ALMA radial velocity are not able to constrain the eccentricity of the orbit. When combined with the astrometry from \textit{Gaia} DR3, the highest likelihood orbits are in the range $\text{ecc}<1$, but the hyperbolic orbits are not excluded, similarly to the case of fitting the ALMA astrometry. 
The astrometry of both instruments was not combined in a single fit, because of a non-negligible difference in astrometry between ALMA and \textit{Gaia} DR3 of about 6mas, which is shown in the middle panel of Fig.~\ref{fig:rwaurb_pos}. This is further discussed in Sect.~\ref{sec:discussion}.

\subsection{Inner disk geometry of RW\,Aur\,A}\label{sec:warp_fit}

\begin{figure}[t]
 \centering
        \includegraphics[width=8.8cm]{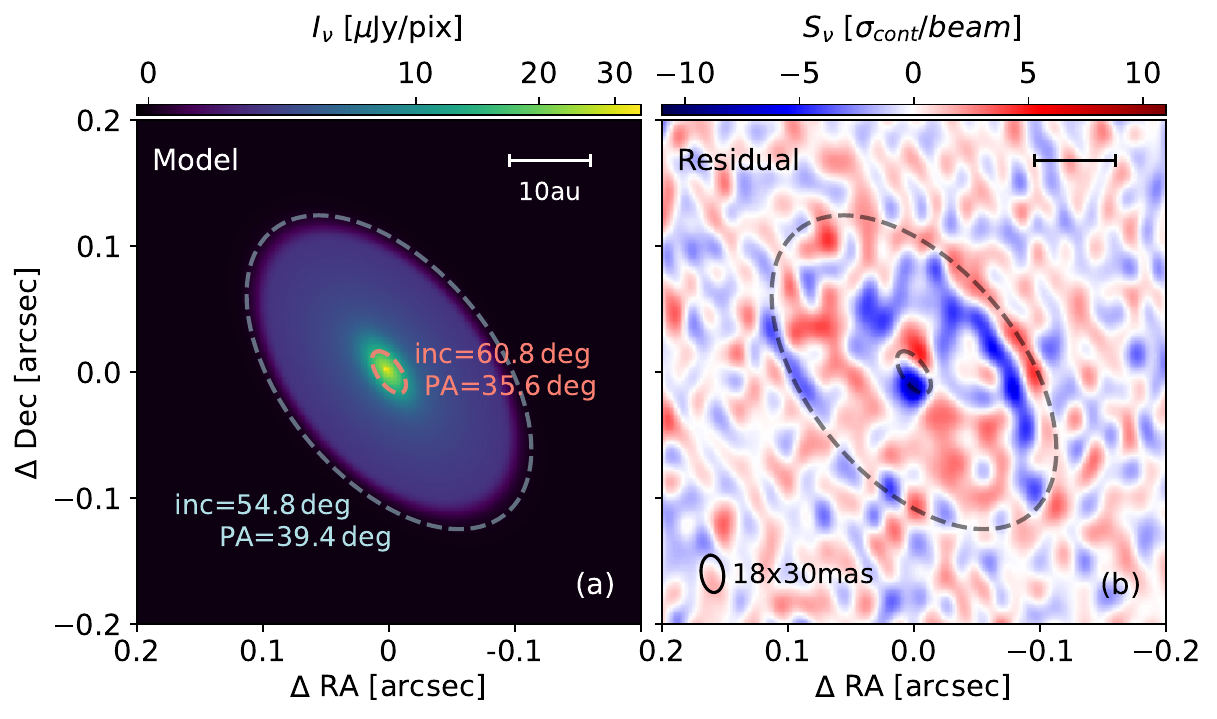}\\
        \vspace{-0.1cm}
   \caption{As in Figure \ref{fig:uvmodel_cont}, but for the warped visibility model described in Sect.~\ref{sec:warp_fit}. The inner disk ellipse in panel \textbf{a)} shows the full width at half maximum size of the central Gaussian, with the geometry from Table~\ref{tab:warp_prop}. }
   \label{fig:uvmodel_cont_11}
\end{figure}

After we subtracted the best visibility model in Sect.~\ref{sec:results_uvmodel}, the residuals in the RW\,Aur\,A inner disk show a dipole-like structure (as shown in panel b) of Fig.~\ref{fig:uvmodel_cont}), which could be due to a slightly different inclination for this region \citep[see compilation of residuals in the Appendix of ][]{andrews2021}. 
On the other hand, the timescale for the latest binary interaction suggested by our orbital fitting was most likely few hundred years ago, which could have misaligned the geometry of the inner and outer disk. Motivated by these results, we ran an additional visibility model for the dust continuum emission in the same way as described in Sect.~\ref{sec:results_uvmodel}. We allowed the central Gaussian component describing the inner disk emission of RW\,Aur\,A to have a different inclination and position angle (inc$_{\text{inn}}$, PA$_{\text{inn}}$) relative to the outer disk (inc$_{\text{out}}$, PA$_{\text{out}}$). The results of the MCMC are shown in Table \ref{tab:warp_prop} and Fig.~\ref{fig:uvmodel_cont_11}.

The central Gaussian finds a higher inclination than the outer disk, with a relative difference of $6.0\pm0.4$\,deg between the inner and outer disk. The Gaussian width at half maximum is 3\,au, indicating the extent of the tentative inner disk warp. When reconstructing an image with the residual visibilities between the observation and the best model, we find that the highest amplitude residual is $5\sigma$, which is lower than for the non-warped disk model. Nonetheless, low-contrast residuals from non-axisymmetric structures are observed throughout the whole disk. The remaining parameters describing the outer disk of RW\,Aur\,A and RW\,Aur\,B remained consistent.

\begin{table}[t]
\centering
\caption{Geometry and inner disk properties of RW\,Aur\,A. }
\begin{tabular}{ c|c|c|c } 
  \hline
  \hline
\noalign{\smallskip}
           & Property      & Best value $\pm 1\sigma$ & unit \\
\noalign{\smallskip}
  \hline
\noalign{\smallskip}
RW\,Aur\,A & inc$_{\text{out}}$ & $54.83 \pm 0.04  $ & deg \\
Geometry   & PA$_{\text{out}}$  & $39.43 \pm 0.05  $ & deg \\
           & inc$_{\text{inn}}$ & $60.82 \pm 0.39  $ & deg \\
           & PA$_{\text{inn}}$  & $35.55 \pm 0.57  $ & deg \\
\noalign{\smallskip}
  \hline
\noalign{\smallskip}
RW\,Aur\,A & $f_{A1}$     & $  12.81 \pm 0.30 $ & $\mu$Jy/pix \\
inner disk & $\sigma_{A1}$& $  16.45 \pm 0.23 $ & mas \\
  \hline
  \hline
\end{tabular}
\vspace{0.1cm}
\tablefoot{ Properties are measured by an MCMC fitting described in Sect.~\ref{sec:warp_fit}. }
\label{tab:warp_prop}  
\end{table}

%%%%%%%%%%%%%%%%%%%%%%%%%%%%%%%%%%%%%%%%%%%%%%%%%%%%%%%%%%%%%%%%%%%%%%%%%%%%%%%%

\subsection{12CO J=2-1 emission}\label{sec:obs_gas}

\begin{figure*}[t]
 \centering
        \includegraphics[width=18cm]{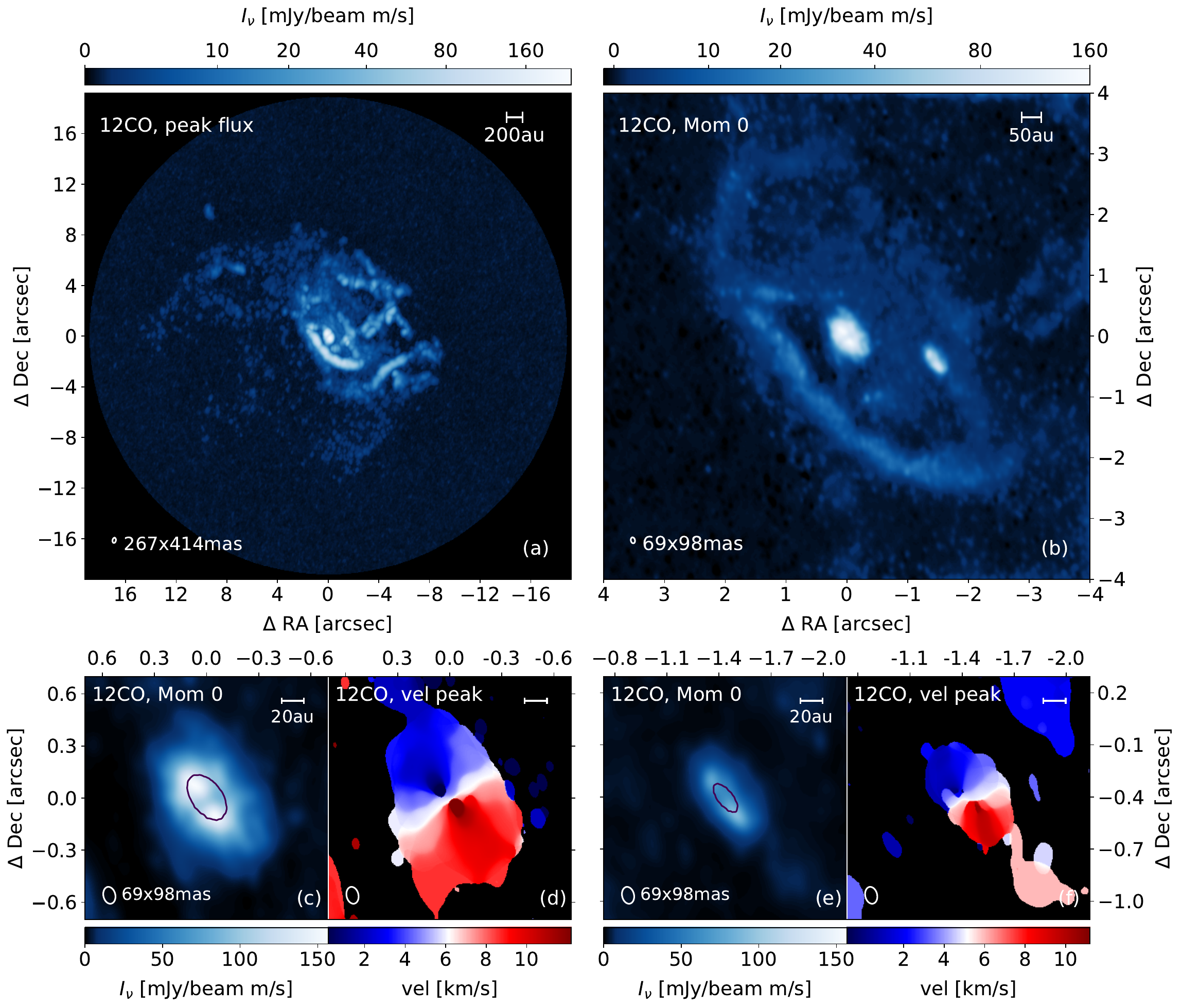}\\
        \vspace{-0.1cm}
   \caption{$^{12}$CO emission images at different spatial scales. \textbf{Panel (a)} shows the peak emission of each pixel, imaged by combining observations SB1, SB2, and SB4. \textbf{Panel (b)} shows the moment 0 image generated by combining all the existing datasets. \textbf{Panels (c)} to \textbf{(f)} were generated by clipping the emission at 1.3\,mJy\,beam$^{-1}$\,m\,s$^{-1}$, and panels \textbf{(c)} and \textbf{(d)} are a zoom into RW\,Aur\,A, while panels \textbf{(e)} and \textbf{(f)} are a zoom into RW\,Aur\,B. \textbf{Panels (d)} and \textbf{(f)} show the velocity of the peak brightness emission. The color bar of the velocity at peak has been chosen so that the central velocity matches the systemic velocity of each disk. The contour in panels \textbf{(c)} and \textbf{(e)} show the $5\sigma$ dust continuum contours.}
   \label{fig:gallery_co}
\end{figure*}

Our $^{12}$CO observation redetects the main spiral arc reported in \citet{Cabrit2006}. The increased sensitivity also allowed us to connect the clumps of emission detected in \citet{rodriguez2018}, previously called $\alpha$, $\beta$, $\gamma$, and $\delta$, into an intricate system of tidal arcs, filaments, and an extended diffuse background emission, which extends farther than 2000\,au in projected distance from RW\,Aur\,A, as shown in panel (a) in Fig.~\ref{fig:gallery_co}. 
This panel also shows the whole field of view of the SB $^{12}$CO image, which combines the observations with short baseline configurations (SB1, SB2, and SB4), thus maximizing the sensitivity over extended spatial scales. 
The LB $^{12}$CO image is shown in panel (b), and its increased angular resolution allowed us to resolve the disk emission and kinematics, shown in panels (c) to (f). 
Component RW\,Aur\,C, proposed in \citet{rodriguez2018}, is not resolved into a disk-like emission or coherent rotating structure. As the LB1 observation has a lower frequency resolution than the other observations, the LB $^{12}$CO image is limited to a velocity resolution of 1.3\,km\,s$^{-1}$, which hides kinematic structures with a smaller velocity amplitude.

We used the package \texttt{eddy} \citep{eddy} to fit the Keplerian rotation of each disk and estimate the mass of the central object. We did not downsample the velocity image pixels, which have a size of 4\,mas. 
The emission was masked using an elliptical mask with a size of 0.4'' and 0.28'' for A and B, respectively, to avoid including non-Keplerian emission that surrounds each object. 
As a result of the compact nature of the sources and the low-frequency resolution, our images cannot distinguish upper and lower emission surfaces. 
Therefore, we fit them with a flat Keplerian disk. The disk center, inclination, and position angle are fixed values from the dust continuum modeling, taking the values obtained for the outer disk. 
The only free parameters for each disk are the stellar mass and the central velocity in the line of sight. For RW\,Aur\,A, we obtain $M_A=1.238\,M_\odot$ and $VLSR_A=6172.44\,$m\,s$^{-1}$, and for RW\,Aur\,B we obtain $M_B=0.995\,M_\odot$ and $VLSR_A=5218.24\,$\,m\,s$^{-1}$. 
The two mass measurements are in agreement with spectroscopic estimates \citep{Herczeg2014, long2019, manara2019}. 
The uncertainty estimate from eddy suggests that the standard deviations are a fraction of a percent for the masses and velocity in the local standard of rest. 
These small uncertainties come from the combination of the very high sensitivity per channel and the low-velocity resolution, which produces a ``channelization'' effect, as described in the second paragraph of Sect.~\ref{sec:rwaurb_orbit}, and shown in panels (c) to (f) from Fig.~\ref{fig:gallery_co}. 
A possible solution of this problem is a forward-modeling of the image cube \citep[e.g.,][]{Izquierdo2021}, or of the visibilities \citep[e.g.,][]{long2021, kurtovic2024b}. For both approaches, a better understanding of the gas disk structure is needed. The possible misalignment of the inner disk was not considered in our fit, as the inner 3\,au are completely contained within the first beam of the LB $^{12}$CO image, and it has therefore little influence on the mass estimate, which is mainly constrained by the outer disk. 

\begin{figure}[t]
 \centering
        \includegraphics[width=8.5cm]{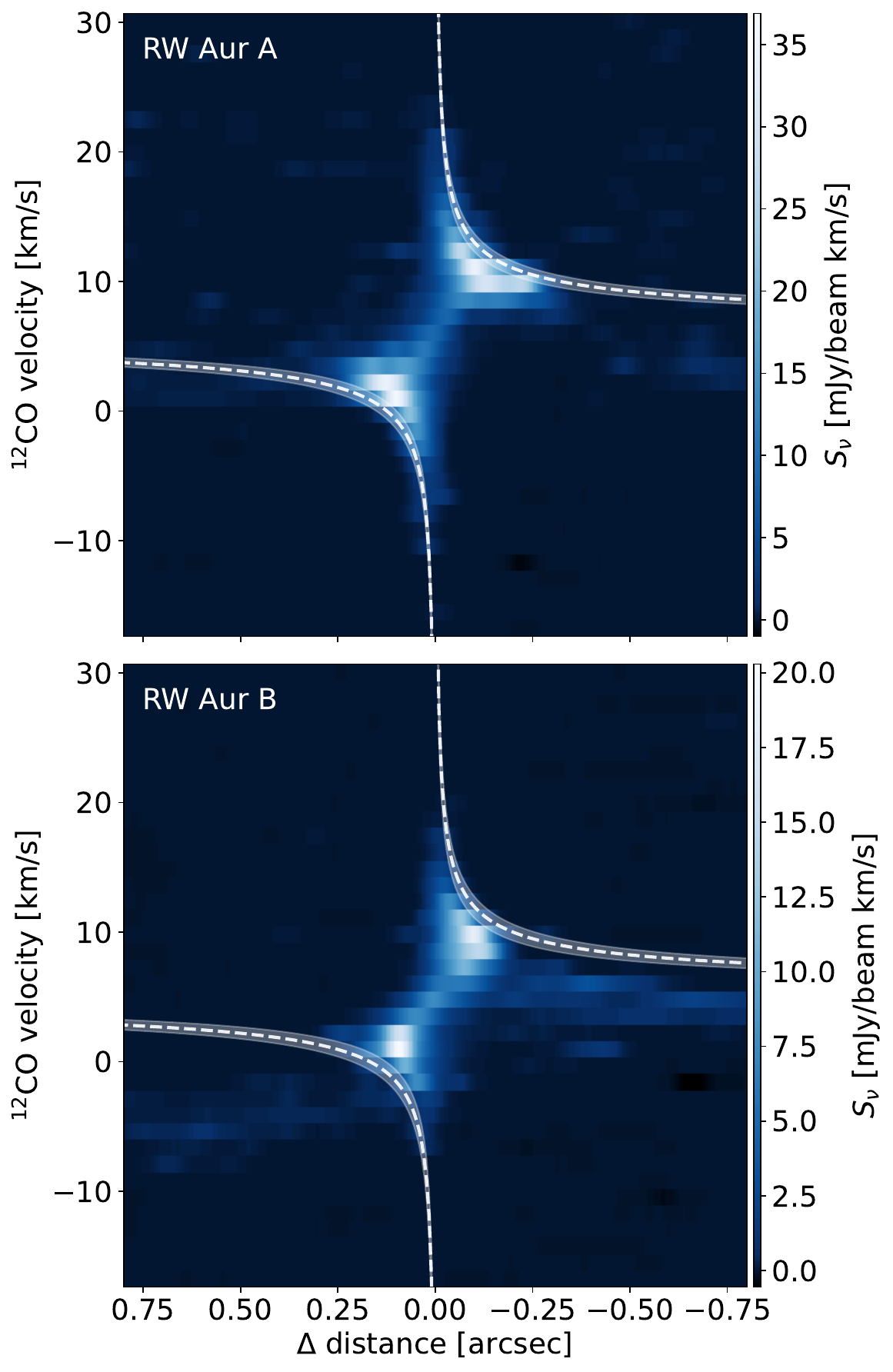}\\
        \vspace{-0.1cm}
   \caption{Position velocity diagram calculated from the LB $^{12}$CO cube, using the inclination and position angle obtained with the dust continuum visibility modeling along the major axis. The dashed rotation curve shows the best solution obtained with eddy, while the shaded region shows the $3\sigma$ confidence region of the rotation curve by changing the stellar mass and disk systemic velocity, as discussed in \ref{sec:obs_gas}.}
   \label{fig:PVdiag}
\end{figure}

The detected emission structures span thousands of astronomical units in spatial scales and about 40\,km\,s$^{-1}$ in velocity. 
Over this extended range, many pixels in the image are part of different $^{12}$CO kinematic structures, such that a single velocity image at peak emission or moment 1 does not represent the kinematic richness of the gas. To alleviate this problem, we separated the channels into the blue- and redshifted emission relative to RW\,Aur\,A and calculated their peak brightness and velocities, as shown in Fig.~\ref{fig:bluered}. 
Most of the redshifted emission is connected to the bright southern arc, while the blueshifted emission has a semi-circle shape, with most of the emission being northwest of RW\,Aur\,A. 
Under the assumption that these structures move farther away from the RW\,Aur disks, the northwestern emission would be closer toward us in the line-of-sight direction, and the redshifted southeastern emission would be farther away, as also proposed by \citet{Cabrit2006}.

The high angular resolution $^{12}$CO moment 0 after continuum subtraction shows that neither of the disks is centrally peaked, as shown in Fig.~\ref{fig:gallery_co}. 
In RW\,Aur\,B such morphology is expected, because there is a cavity in the dust continuum emission. However, it is unexpected in RW\,Aur\,A, where the dust continuum is centrally peaked. In this disk, the morphology of the $^{12}$CO moment 0 could be influenced by oversubtraction of an optically thick dust continuum emission and beam dilution in the central regions \citep[e.g.,][]{wolfer2021}. 
When we measured the location of the peak emission in $^{12}$CO, we obtained a peak at $113\pm{4}$\,mas (about 18\,au) for RW\,Aur\,A and a peak at $93\pm{4}$\,mas (about 15\,au) for RW\,Aur\,B, roughly at the locations of their outer radii in the dust continuum emission. 
As the disks are surrounded by material from the interaction, the definition of the outer radii for the gas is not as simple as in isolated disks. To avoid including emission from the surrounding material, we integrated the flux over an elliptical aperture with a radius of $0.6''$ and $0.45''$ for RW\,Aur\,A and B, respectively. 
Farther than these distances, the flux contribution is dominated by the surrounding material, and not by the disks. 
The radii enclosing the 68\% and 90\% of the flux for RW\,Aur\,A are $R_{A, CO, 68\%}=327\pm25\,$mas and $R_{A, CO, 90\%}=456\pm64\,$mas, and for RW\,Aur\,B, we obtain $R_{B, CO, 68\%}=238\pm29\,$mas and $R_{B, CO, 90\%}=343\pm60\,$mas. 
These radii are smaller than those measured by \citet{rota2022}. The main difference is the cutoff radius for the integrated flux. When compared to the R90\% of the continuum radius shown in Table~\ref{tab:disk_prop}), both disks have a gas-to-dust size ratio of 3.7, which is consistent with dust evolution by radial drift \citep{trapman2019, zagaria2021}. 
However, this value should be considered as an upper limit, because the gas radius was measured from a beam-convolved $^{12}$CO profile. 
It is relevant to note that the $R_{90\%}$ could still be influenced by perturbed material at the outer edge of the disks, and the disk radii where the material follows Keplerian motion could be smaller.

We combined all the compact antenna observations (SB1, SB2, SB3, and SB4) to produce images for $^{13}$CO and C$^{18}$O, using the same imaging parameters as the $^{12}$CO. We calculated their velocity-integrated maps after clipping by $3\sigma$, and the results are shown in Fig.~\ref{fig:co_iso}. In the $^{13}$CO image, the sensitivity is high enough to detect emission from the southern arc, but in C$^{18}$O, we only have a detection of the RW\,Aur\,A disk. The angular resolutions of these images are similar to that of the SB $^{12}$CO image, and therefore we cannot resolve the cavities or structures, if they exist.

%%%%%%%%%%%%%%%%%%%%%%%%%%%%%%%%%%%%%%%%%%%%%%%%%%%%%%%%%%%%%%%%%%%%%%%

\section{Discussion} \label{sec:discussion}

\subsection{ Testing the bound/unbound nature of the RW\,Aur system} \label{sec:discussion:orbit}

The RW\,Aur binary system has two families of solutions that are consistent with the current astrometry: Bound orbit solutions, most of them in the eccentricity range between $\text{ecc}\in[0.5\sim0.95]$, and unbound solutions with eccentricities extending up to $\text{ecc}\approx 3$, as shown in Fig.~\ref{fig:orbit_prob_dens}, and Fig.~\ref{app:fig:orbit_prob_dens}. 
There is a clear likelihood difference among the allowed orbits, with a group of elliptical orbits having a higher likelihood of describing the data than any of the hyperbolic orbits. We compare the solutions with the Akaike information criterion \citep[AIC,][]{akaike1974}, which quantifies the loss of information for a given model. When comparing the best $99.85$ percentile of elliptical and hyperbolic models, we find that the best hyperbolic models are 0.28 times as likely as the best elliptical models to be the solution that minimizes the information loss. Considering this probability in units of $\sigma$, there is a $\sim1.1\sigma$ of confidence that elliptical orbits are preferred over hyperbolic orbits. This comparison only takes into account the likelihood from the astrometric fit, and does not consider as a prior the morphology of the $^{12}$CO emission, which favors elliptical orbits to explain the extended emission.

When the elliptical and hyperbolic solutions are extended into the next decade, we predict that a single astrometric measurement is unlikely to distinguish between the bound or unbound solutions until after 2037 (see Fig.~\ref{app:fig:future}), and thus, several measurements will be needed to continue to improve the orbit prediction. 
In radial velocity, both model families show a very small variation from the current measurement ($<0.1\,$km\,s$^{-1}$) over the next two decades, which means that accurate astrometry will be the determining measurement to distinguish the nature of the RW\,Aur orbit. 
Because hyperbolic and elliptical orbits show distinct distributions of $inc$, $\Omega$, and $\omega$, a better constraint on any of these elements would also contribute to distinguishing the orbital eccentricity.

Future observations will improve our estimates for the distance at last periastron, which currently range between tens of au to about 600\,au. When considering $t_{peri}$ to be older than the time of the ALMA observations, we find that $t_{peri}$ is most likely between 200\,yr to 500\,yr ago, although more recent periastrons are not been excluded with the current astrometry.

\subsection{Evidence of interaction in the 12CO emission} 

Previous simulations of a close encounter in RW\,Aur have shown that an interaction could excite tidally stripped arcs of material, such as the one observed in the $^{12}$CO emission \citep{Dai2015}. 
The additional clumps of $^{12}$CO detected by \citet{rodriguez2018} that we redetected at higher sensitivity (see panel (a) in Fig.~\ref{fig:gallery_co}) could have been produced as tidal arcs in previous interactions of this bound system, which would be consistent with the elliptical orbit solutions.

Tidal interactions are expected to truncate the disks sizes, and simulations and analytical studies both found that disks in multiple stellar systems are truncated to a fraction of the binary separation \citep[typically 0.3-0.5, e.g.,][]{Clarke1993, Pichardo2005, Harris2012, Rosotti2014, manara2019, zurlo2020, zagaria2023}. 
In RW\,Aur, about $18\%$ of the elliptical orbit solutions suggests that the distance at the closest interaction could be smaller than the measured gas extent (see Fig.~\ref{fig:orbit_prob_dens}), which contradicts the truncation scenario. Additional observations are needed for a more robust constraint on the size of the Keplerian disk around each star, and thus, to properly test their dynamical truncation.

When the orbital plane of the binary stars is not aligned with the plane of their disks, a warp can be excited during a close interaction \citep[e.g.,][]{papaloizou1995, Cuello2019, Nealon2020, gonzalez2020}. As the gravitational influence of the perturber decreases over time after periastron, the warp can smooth out toward coplanarity \citep[e.g.,][]{picogna2014, martin_reb2019, Rowther2022}, and change the disk plane during this process. 
When we assume that the binary orbit is elliptical, then the circumstellar material will change their relative disk-binary inclination with every interaction, and thus, the next tidally stripped arc of material could be ejected in a different direction. This effect, added to possible temperature differences due to stellar illumination \citep[e.g.,][]{Weber2023}, are tentative explanations for the $^{12}$CO emission structure.

When the LB1 observation is included, both disks are spatially resolved in $^{12}$CO, allowing us to analyze their Keplerian rotation. 
Due to the low-frequency resolution, we are unable to confirm or exclude warped or tidally induced velocity structures, as has been observed in other systems \citep[e.g.,][]{kurtovic2018, Mayama2018}. Additional high angular resolution observations toward RW\,Aur at a higher frequency resolution could explore this kinematic aspect of the interaction, and also allow us to estimate each stellar mass more robustly.

Neither disk is centrally peaked in the $^{12}$CO integrated intensity map, as shown in panels c and e in Fig.~\ref{fig:gallery_co}. In RW\,Aur\,B, a cavity is also observed in the dust continuum emission. In RW\,Aur\,A, however, the dust continuum emission is centrally peaked, and a cavity in the gas emission is accordingly puzzling. As the peak emission of $^{12}$CO in RW\,Aur\,A is detected at a similar distance as the outer disk continuum radius, an optically thick dust continuum emission could result in a continuum oversubtraction, thus contributing to the observed cavity. 
However, the role of the accretion events, inner disk misalignment, and the possible asymmetry in the morphology of $^{12}$CO remains an open question. Observations at similar angular resolution to LB1 but at higher frequency resolution should be able to characterize the disk gas morphology, which would also allow us to determine the orientation of the inner disk from kinematics.

\subsection{Disks structure in dust continuum emission} \label{sec:discussion:cont_structure}

We confirm the compact nature of the dust continuum disk sizes, as was also observed by \citet{rodriguez2018} and \citet{manara2019}. 
Although the RW\,Aur\,A dust continuum image (as reconstructed by CLEAN) seems to be a featureless disk, the residuals from our visibility modeling show that it is rich in low-contrast small-scale structure. A geometrically flat disk model describes most of the emission of the disk, and it only leaves a strong structured residual in the inner disk region, suggesting a different inclination for inner and outer disk. 
From an additional visibility modeling, in which we allowed the inner disk region emission to have a different inclination, we find a difference of $6.0\pm0.4\,$deg between the inner ($<3$\,au) and outer disk ($>3$\,au) and a difference in PA of $4.0\pm0.6\,$deg. This tilt could have originated in the last close encounter between the two stars. 
Additional structure is observed in the residual map after subtracting the model with a misaligned inner disk (see Fig.~\ref{fig:uvmodel_cont_11}), suggesting that the inner disk of RW\,Aur\,A could have an azimuthally asymmetric structure, as was observed in the inner disk of other systems \citep[e.g., ][]{ganci2024}. Higher angular resolution observations are needed to confirm this geometry and morphology, and this makes RW\,Aur\,A an ideal candidate for observations with the near-infrared interferometric capabilities of the Very Large Telescope. 

The misalignment we observe in RW\,Aur\,A is likely to be smaller than the initial tilt induced by RW\,Aur\,B during the last periastron. With SPH simulations, \citet{Rowther2022} showed that the timescale for smoothing out a misalignment can be as short as a few orbits of the outer disk edge, which is consistent with the time since last periastron recovered with our orbital fittings. \citet{Rowther2022} also showed that the difference in the velocity field between the inner and outer warped disk can produce work and dissipate energy as heat. Under the assumption of optically thick dust continuum emission, the misalignment might contribute to the high brightness temperature observed in the midplane of RW\,Aur\,A. 

More generally, the evolution of a warped structure over time depends on the viscosity of a disk, and in the low-viscosity regime, a warp can have a wave-like behavior \citep{lubow2000, martin_reb2019, dullemond2022, kimmig2024}. \citet{Koutoulaki2019} explored the role of a misalignment between inner and outer disk in the dimming events that have been observed since 2010, showing that for a disk with $T=30$\,K at $10$\,au and a bending-wave starting at $58$\,au, the warp would take about 520\,yr to reach the inner disk. The last periastron passage derived from the elliptical orbits with the ALMA astrometry is consistent in order of magnitude with this estimate ($295^{+21}_{-74}$\,yr from the beginning of 2024), and is also consistent with the last periastron of the hyperbolic orbits. Additionally, our findings indicate that the inner 3\,au show a higher inclination than the outer $3$\,au disk, which supports the hypothesis that material in a misaligned inner disk contributes to the dimming events \citep{facchini2016}.

Considering the results of our elliptical and hyperbolic orbital fitting, we can speculate about the origin of the dust structures in the disk of RW\,Aur\,B. The distribution of elliptical solutions for the binary periastron distance accumulates at small radii, with a peak likelihood comparable to the gas radii of RW\,Aur\,A, with orbital solutions as small as their dust size. These orbits are not consistent with expected disk truncation from close interactions, and no strong constraints over the minimum interaction distance can be set with the current astrometric data. These small $d_{peri}$ do not exclude the possibility of a disk collision, although it should be noted that we cannot infer the disk size before the periastron, and similarly, we also find solutions for the periastron that are larger than the emission radii and are more consistent with dynamical truncation. 

The close encounter, elliptical or hyperbolic, would have induced a warp in RW\,Aur\,A, while RW\,Aur\,B could have captured some material into its own disk \citep[even if the periastron is larger than the disk size,][]{Clarke1993}, as was observed in SPH simulations of fly-by encounters \citep[e.g.,][]{Dai2015, Cuello2019, gonzalez2020}. 
The lower surface density of captured grains could translate into a lower optical depth, which is consistent with the low brightness temperature of RW\,Aur\,B. The captured material scenario could be tested with dedicated simulations using the recovered orbital parameters and the observed geometry of the two disks.

Our speculative interpretation of the disk structures would benefit from additional observations at high angular resolution. A follow-up with ALMA starting from cycle 11 would increase the time baseline of the ALMA astrometry by a factor of four compared to this work, and it would allow a more robust determination of the time and distance at periastron, for either the bound or unbound solutions. Similarly, a follow-up in longer millimeter wavelengths will enable accurate measurements of the spectral index and optical depth, which could provide an additional test of our explanations for the origin of the substructures by better constraining the dust density and temperature distribution of each disk.

\subsection{Orientation of the orbit, disks, and jet emission}

The geometry of each disk is constrained based on our dust continuum observations. Additionally, we spatially resolve the blue- and redshifted sides in the $^{12}$CO kinematics (see Fig.~\ref{fig:gallery_co}). Our angular and frequency resolution are not high enough to differentiate between the upper and lower emission layers of the disk, and consequently, we are unable to determine the orientation of the angular momentum vector for each of them using only the CO emission. However, RW\,Aur\,A has a well-studied jet with a redshifted component oriented in the northwest direction \citep{Dougados2000}. Assuming that the disk is close to perpendicular to the jet \citep[as it has been observed in other young sources, e.g., ][]{burrows1996, floresrivera2023}, we can infer from the rotation map shown in Fig.~\ref{fig:gallery_co} that the angular momentum vector of RW\,Aur\,A points in the northwest direction as well, which results in a clockwise rotation of the disk in the sky plane. 

In a coordinate system centered on RW\,Aur\,A, RW\,Aur\,B also orbits in clockwise orientation. When we compare the inclination and position angle of RW\,Aur\,A disk, using the orientation implied by the jet emission, with the inclination and longitude of the ascending node of the orbital solutions from the MCMC, we obtain a disk-binary misalignment of ${27.1}\pm26.7$\,deg, where the errors represent the $1\sigma$ dispersion around the best value. The last interaction was most likely at a prograde orientation relative to RW\,Aur\,A, although more polar interactions up to a misalignment of $\approx70$\,deg are not ruled out. We exclude the scenario of a strong retrograde interaction.

\subsection{Origin of the close interactions}

There are a few speculative scenarios for the origin of the close interactions. For example, if the last interaction was hyperbolic in nature, then the stars would be gravitationally unbound before and after the interaction, also known as a fly-by. Even though this scenario is not excluded with the current astrometric measurements, it is more challenging to reconcile with the multiple filaments and features observed in the $^{12}$CO emission.

We can also consider the hypothesis that the RW Aur binary has only recently been induced into a highly eccentric elliptical orbit. This could have occurred through stellar capture from an initially unbound interaction, as proposed by \citet{rodriguez2018}, or alternatively, RW\,Aur\,A and B could have been induced into a highly eccentric elliptical orbit through an interaction with a third body. The dissolution of triple stellar systems commonly results in the formation of a single and a binary stellar system \citep[e.g.,][]{Toonen2022}, and interactions with external gravitational potentials (e.g., a third companion) can change the eccentricity of the bound binary \citep[e.g.,][]{Monaghan1976, Stone2019, Ginat2021}.

By analyzing \textit{Gaia} DR3 proper motion and parallax, \citet{Shuai2022} found that \textit{Gaia} DR3 156431440590447744 could have had a closest approach at a distance of $13.2\pm2.5$\,kau with RW\,Aur about $6\cdot10^3$\,yr ago. Although this large distance makes an interaction unlikely, both RW Aur A and RW Aur B have \textit{Gaia} RUWE values over 1.4 (16 and 1.5 respectively), and thus their parallax and proper motion should be reanalyzed in future works considering their binarity and variability. Observations over a longer time baseline with high-precision parallax measurements could test this third-companion hypothesis and potentially improve our constraints on the dynamical history of the RW\,Aur binary.

\subsection{Astrometry with ALMA}\label{sec:discussion:astrometry_alma}

Due to their compact emitting surface, stars are usually undetected in millimeter-wavelength observations. As stars cannot be directly observed with instruments such as ALMA, it is challenging to find a reference for high-precision astrometry. When binary disks are detected, as in RW\,Aur, high-precision astrometry can be performed in a relative coordinate system by fixing the origin on the center of one of the disks and calculating their relative separation. 

The center of each disk can be recovered with MCMC approaches such as parametric visibility modeling. To translate disk astrometry into stellar astrometry, we need the additional assumption that the stars are in the center of the disks. Even though this is a safe assumption for most of the study cases, ALMA is sensitive to very low contrast asymmetries and eccentricity structures \citep[e.g.,][]{andrews2021, kurtovic2022}, which could shift a disk center by a few milliarcseconds. Thus, any attempt to recover stellar astrometry from modeling the disk's position must use a model that describes the emission morphology as well as possible. 

In RW\,Aur\,A, our parametric model for the dust continuum did not consider azimuthal asymmetries at any radius, which left low-contrast structure in the residual image (see Fig.~\ref{fig:uvmodel_cont_11}). 
These residuals are problematic, because an asymmetry in the inner disk region could slightly shift the recovered center of the disk. We considered an azimuthally symmetric disk brightness distribution with a flux of $L_s$, with a compact asymmetry in the inner disk region of brightness $L_a$ and a distance from the geometric center $r_a$. The shift of the disk light center $r_{lc}$ from the geometric center will be $r_{lc} = r_a L_a / (L_s+L_a)$. For an asymmetry at $r_a\approx1$\,au, the shift of RW\,Aur\,A would be $r_{lc}\approx 0.19L_a$ with $L_a$ in mJy and $r_{lc}$ in milliarcseconds. Thus, over the span of a single orbit, an asymmetry of $L_a\approx2.5\,$mJy could shift the light-center as much as 1\,mas, which would modify the recovered disk center compared to an axisymmetric model. This effect is a possible explanation for the difference in relative astrometry between the observations SB1-LB1 and LB1-SB4, which do not show the same relative movement between the fitted disk centers even though they have a similar time baseline. Longer time baselines for astrometry should be less sensitive to asymmetries rotation.

In addition to carefully describing the disk morphology, additional considerations should be taken when recovering the relative disk astrometry from ALMA data. For example, the visibility weights of observations from different ALMA cycles should be standardized with tools such as \texttt{statwt} from \texttt{CASA} before they are compared with an MCMC-based approach. 
As the flux calibration of ALMA can vary by up to 10\%, a flux-scaling factor should always be fit as part of the analysis process. Finally, if astrometry is the goal of an observation, the S/N should be high enough to be self-calibrated by itself, without the need of combining it with another observation that was taken at a different epoch.

For RW\,Aur, all of our observations were taken at almost the same frequency range (see Table~\ref{tab:obs_log}), and we therefore assumed that the emission morphology was the same for every observation (neglecting the possible changes due to inner disk rotation discussed in the previous paragraphs). Attempts to obtain relative astrometry with parametric models from observations taken at different wavelengths should consider the wavelength dependence of the emission morphology, as structures can change in optical depth, and thus different regions of the disks do not necessarily have the same spectral index. In this scenario, a single flux-scaling factor will not work properly.

The relative binary motion from ALMA has conflicting values when compared to that from \textit{Gaia} DR3, which prevents a simultaneous fit. A likely explanation for their difference are the systematics from \textit{Gaia} when analyzing stars with circumstellar material and variable brightness, quantified by the high RUWE values \citep{Fitton2022}, and the possible shifts due to asymmetries in the inner disk region. The RUWEs of RW\,Aur\,A and B are respectively 16 and 1.5, respectively, which is above the robustness threshold considered by \citet{fabricius2021}. The stellar occultation events in RW\,Aur\,A during the observing period of \textit{Gaia} DR3 might also have introduced systematic errors in the AB relative positions, as well as the ejection of jet knots every 2-6 years \citep{takami2020}. 
Additional observations with ALMA and a reanalysis of the \textit{Gaia} data should alleviate this issue.

When a high S/N observation with ALMA antenna configuration C-10 is analyzed, we obtain binary relative distances with an accuracy comparable to that of \textit{Gaia} DR3. Thus, ALMA observations arise as an alternative for studying binary motion in young star-forming regions in which optical wavelengths are completely extincted by cloud contamination. Dedicated continuum observations at high angular resolution of regions such as Ophiuchus would allow us to study the impact of binarity and interaction in the very early stages of planet and star formation.

Another relevant approximation taken in this work was to consider the stars as point sources that contained all of the mass of the system, thus neglecting the mass distribution and contribution of their disks and surroundings. Considering the compact nature of the disks, and that the disk mass fraction is typically considered to be $<0.05M_\star$, we find that a possible contribution from the disk masses would be within the uncertainty of the total system mass. However, the validity of this approximation should be considered separately for each case-study.

%%%%%%%%%%%%%%%%%%%%%%%%%%%%%%%%%%%%%%%%%%%%%%%%%%%%%%%%%%%%%%%%%%%%%%%

\section{Conclusions} \label{sec:conclusions}

We analyzed the 1.3\,mm emission of the RW\,Aur system, as observed by ALMA over a span of two years, with angular scales ranging from the ACA-7m array to the ALMA C-10 antenna configuration. We resolved the disks in continuum and in $^{12}$CO emission, and confirmed their compact nature. When analyzed in the visibility plane, RW\,Aur\,A shows evidence of low-contrast non-axisymmetric structures, and the inner 3\,au of the disk are tentatively misaligned by 6\,deg relative to the outer disk. RW\,Aur\,B is well described by a single ring that peaks at 6\,au in the dust continuum emission and shows a very low brightness temperature compared to RW\,Aur\,A.

Our $^{12}$CO observations at high angular resolution allowed us to constrain the binary mass under the assumption of Keplerian rotation for their disks. By analyzing the relative separation of the disks as a function of time and combining ALMA with historical astrometry and stellar mass, we constrained the allowed orbital parameter space for the RW\,Aur binary. We find that the most likely solutions are in the elliptical regime (gravitationally bound), indicating to a periastron epoch about 295\,yr ago, but hyperbolic solutions (gravitationally unbound) are not yet excluded.

When fitting elliptical orbits, both models with the ALMA astrometry or \textit{Gaia} DR3 astrometry find consistent results, with highly eccentric gravitationally bound orbits. The hyperbolic models become more likely when approaching $\text{ecc}=1$ from above. Overall, elliptical and hyperbolic solutions agree that the last periastron did not occur more than $\approx500$\,yr ago, which is very recent in astronomical times. 
The tentative warp of RW\,Aur\,A and the brightness temperature structure of both disks are consistent with this very close interaction, whether bound or unbound. Additional observations are needed to confirm or reject this hypothesis, and should mainly be focused on obtaining a better constraint of the distance at last periastron and the physical properties of each disk.

The gas emission of RW\,Aur is resolved into an intricate system of extended low surface brightness emission and filamentary structures, which could be evidence of several close interactions, as previously proposed by \citet{rodriguez2018}, thus supporting the elliptical solutions over the hyperbolic case. Due to the limited frequency resolution of our observations, we are unable to confirm warped structures in the gas emission. 

Multiple-epoch observations of binary systems with ALMA are a viable alternative to recover the stellar orbital parameters, which are crucial for understanding the impact of multiplicity on the planet formation potential of each disk. Even though careful visibility modeling is needed to recover robust disk astrometry, ALMA observations can be used as an alternative to \textit{Gaia} in systems in which optical emission is entirely extinct, thus positioning ALMA as an ideal tool to follow young binary disks over the long term.

\section*{Acknowledgments}

N.K. and P.P. acknowledges support provided by the Alexander von Humboldt Foundation in the framework of the Sofja Kovalevskaja Award endowed by the Federal Ministry of Education and Research. 
S.F. is funded by the European Union under the European Union's Horizon Europe Research \& Innovation Programme 101076613 (UNVEIL). CFM is funded by the European Union (ERC, WANDA, 101039452). Views and opinions expressed are however those of the author(s) only and do not necessarily reflect those of the European Union or the European Research Council Executive Agency. Neither the European Union nor the granting authority can be held responsible for them. This project has received funding from the European Research Council (ERC) under the European Union's Horizon 2020 research and innovation programme (PROTOPLANETS, grant agreement No.~101002188). 
RB is acknowledges support from the Royal Society in the form of a University Research Fellowship. 
CNK aknowledges funding from the DFG research group FOR 2634 ``Planet Formation Witnesses and Probes: Transition Disks'' under grant DU 414/23-2 and KL 650/29-1, 650/29-2, 650/30-1. 
This paper makes use of the following ALMA data: \\ ADS/JAO.ALMA\#2015.1.01506.S, \\ ADS/JAO.ALMA\#2016.1.00877.S, \\ ADS/JAO.ALMA\#2016.1.01164.S, \\ ADS/JAO.ALMA\#2017.1.01631.S, \\ADS/JAO.ALMA\#2018.1.00973.S. \\ ALMA is a partnership of ESO (representing its member states), NSF (USA) and NINS (Japan), together with NRC (Canada), MOST and ASIAA (Taiwan), and KASI (Republic of Korea), in cooperation with the Republic of Chile. The Joint ALMA Observatory is operated by ESO, AUI/NRAO and NAOJ.

\bibliographystyle{aa.bst}
\bibliography{ms.bib}

%%%%%%%%%%%%%%%%%%%%%%%%%%%%%%%%%%%%%%%%%%%%%%%%%%%%%%%%%%%%%%%%%%%%%%%%%%%%%%%%
%                                   Appendix                                   %
%%%%%%%%%%%%%%%%%%%%%%%%%%%%%%%%%%%%%%%%%%%%%%%%%%%%%%%%%%%%%%%%%%%%%%%%%%%%%%%%

\onecolumn

\begin{appendix}

\section{Dust continuum parametric models} \label{sec:annex_uv-model}

The dust continuum emission of RW\,Aur\,A is described with a point source, a centrally peaked Gaussian and a tapered power law function. The tapered power law is shown in Eq.~\ref{eq:tapered}, where the free parameters are the flux amplitude $f_{A2}$, a critical radius $R_{A2}$, and two exponents ($\alpha, \beta$). The best values and $1\sigma$ uncertainty obtained from the visibility model are shown in Table~\ref{tab:uv_mcmc}

\begin{equation}
    TP(r, f_{A2}, R_{A2}, \alpha, \beta) = f_{A2} \left(\frac{r}{R_{A2}}\right)^\alpha 
            \left( 1 - \exp\left[\left(\frac{r}{R_{A2}}\right)^\beta\right] \right)
    \label{eq:tapered}
\end{equation}

\begin{figure*}[t]
 \centering
        \includegraphics[width=18cm]{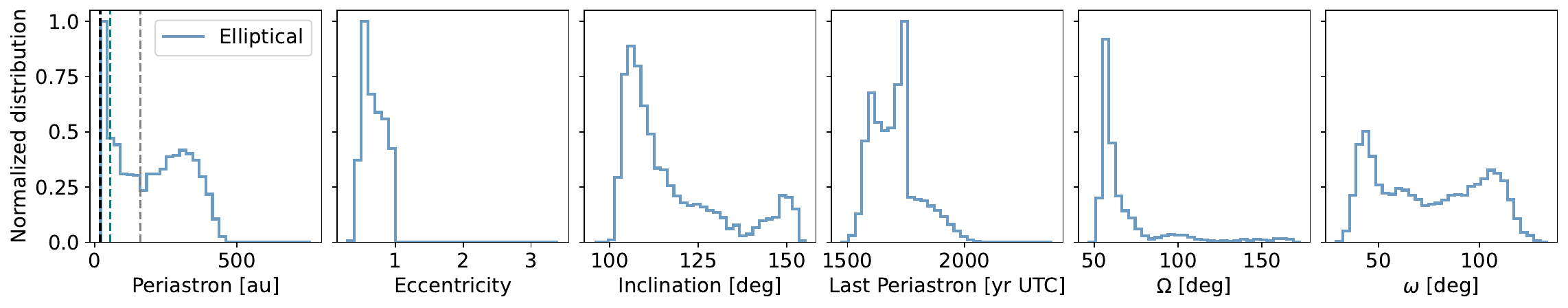}\\
        \includegraphics[width=18cm]{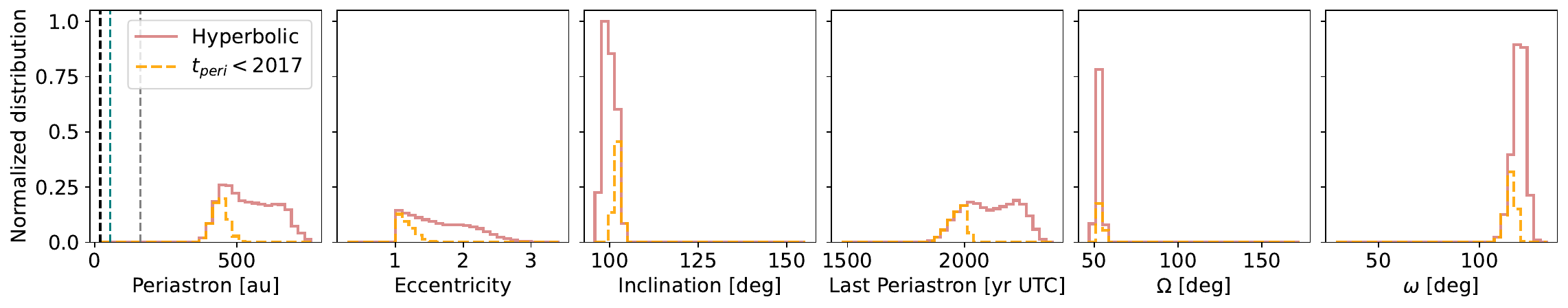}\\
        \vspace{-0.1cm}
   \caption{ Same as Fig.~\ref{fig:orbit_prob_dens}, but showing the solutions with elliptical orbits in the upper panels, and with hyperbolic orbits in the lower panels. Hyperbolic solutions where the periastron epoch was before the first ALMA observation are shown with orange dashed line. Vertical lines in the left panel are as in Fig.~\ref{fig:orbit_prob_dens}.}
   \label{app:fig:orbit_prob_dens}
\end{figure*}

\begin{figure*}[t]
 \centering
        \includegraphics[width=18cm]{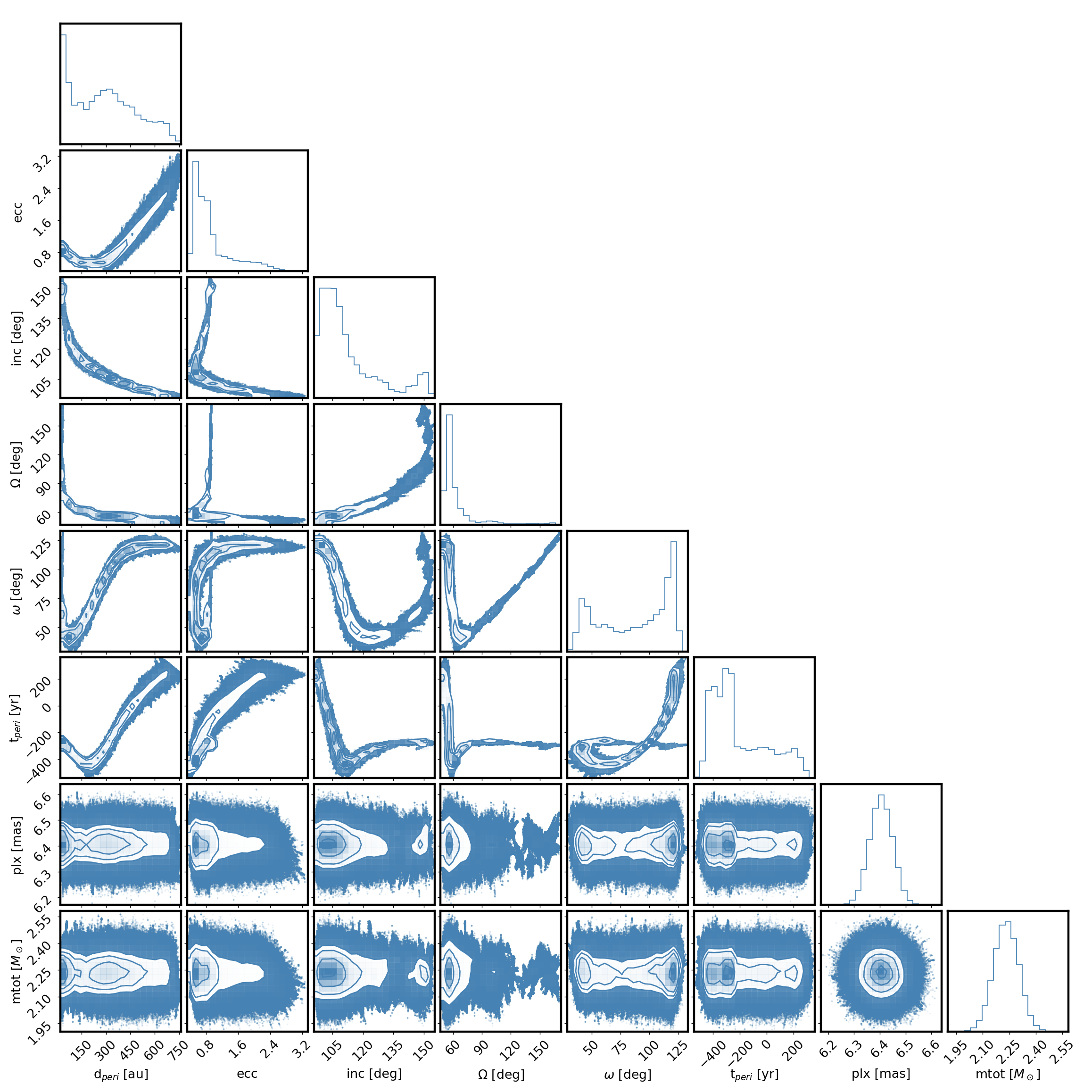}\\
        \vspace{-0.1cm}
   \caption{Corner plot of all orbital solutions. }
   \label{app:fig:corner_all}
\end{figure*}

\begin{figure*}[t]
 \centering
        \includegraphics[width=18cm]{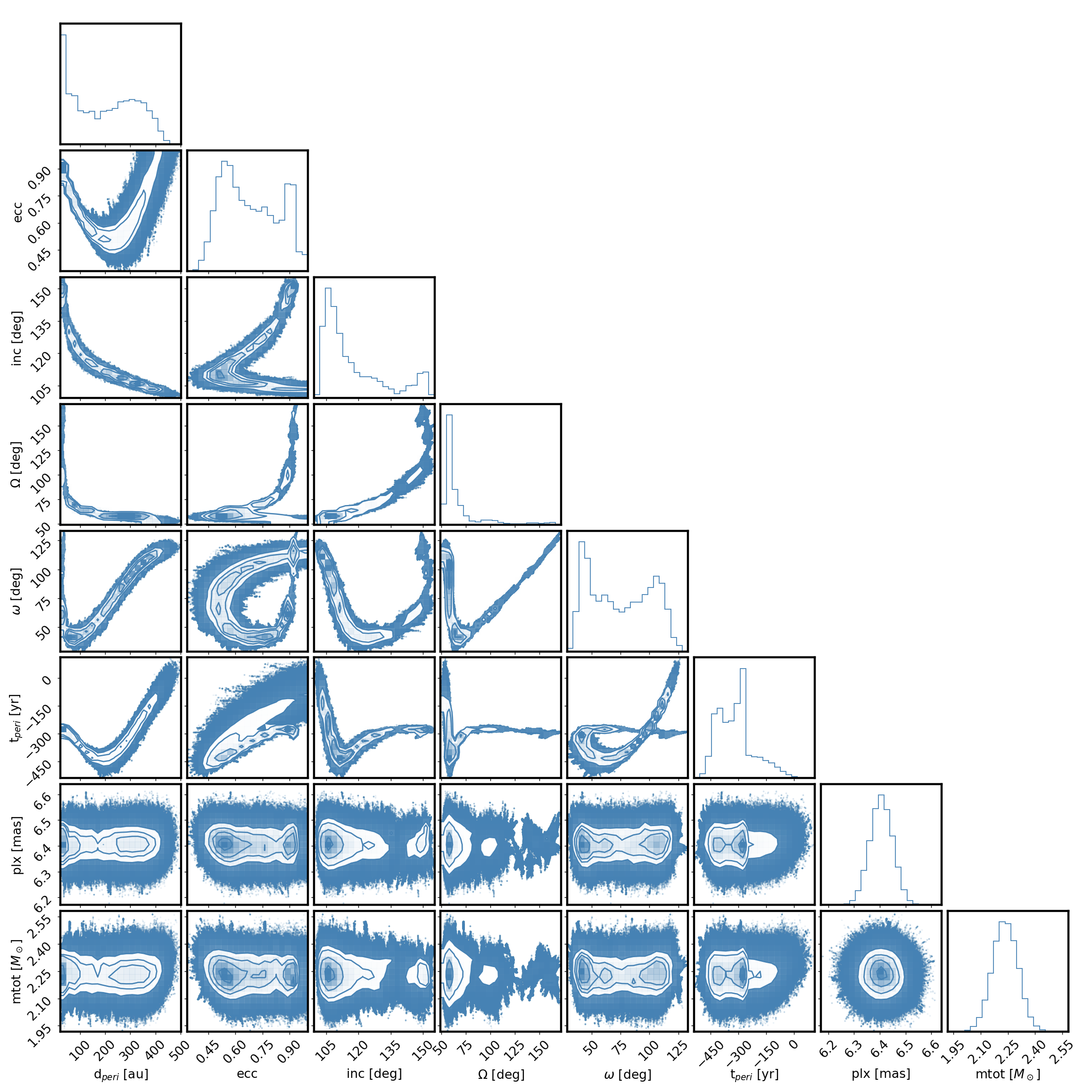}\\
        \vspace{-0.1cm}
   \caption{As Fig.~\ref{app:fig:corner_all}, but only for elliptical solutions.}
   \label{app:fig:corner_ell}
\end{figure*}

\begin{figure*}[t]
 \centering
        \includegraphics[width=18cm]{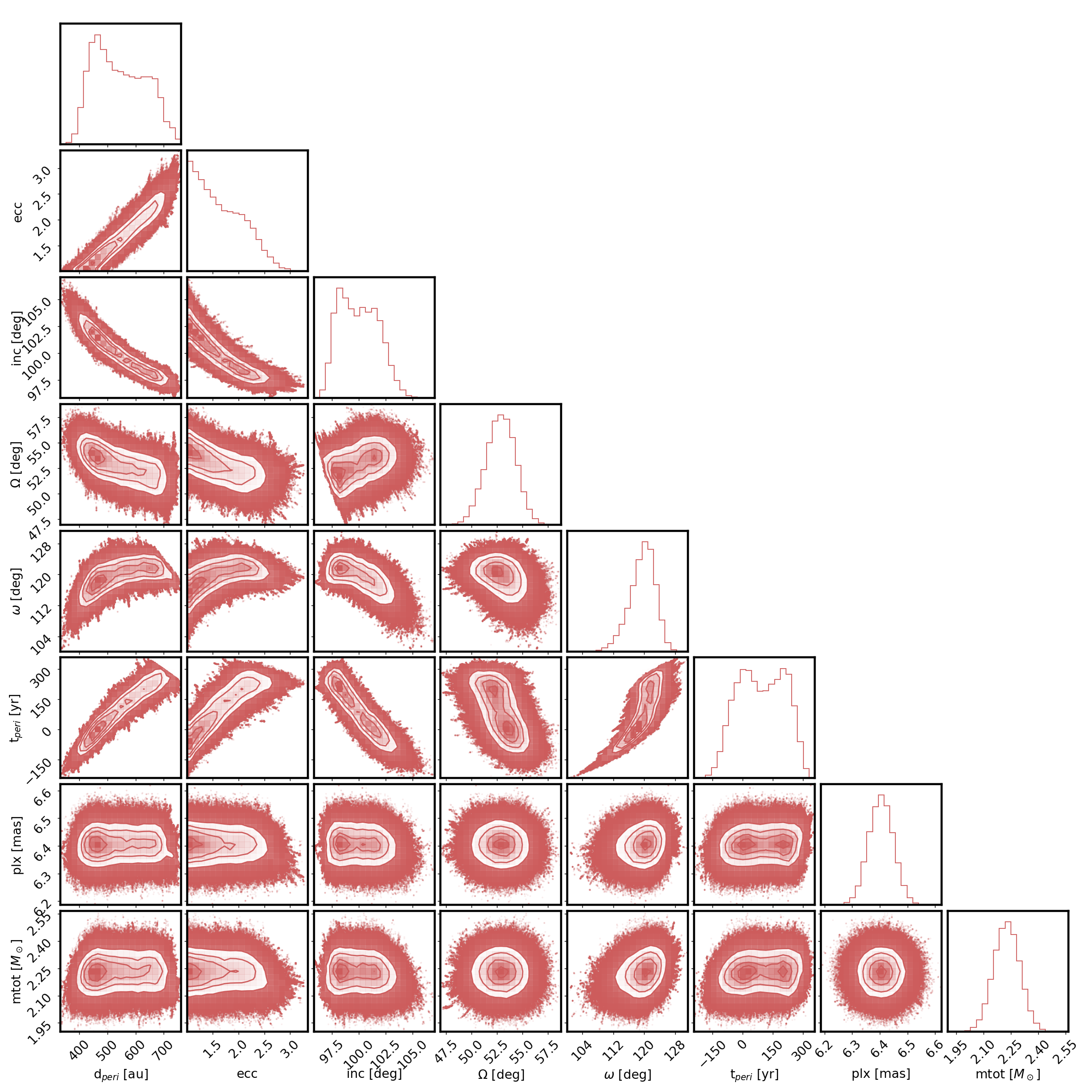}\\
        \vspace{-0.1cm}
   \caption{As Fig.~\ref{app:fig:corner_all}, but only for hyperbolic solutions.}
   \label{app:fig:corner_hyp}
\end{figure*}

\begin{table}[t]
\centering
\caption{ \centering Visibility modeling results for the dust continuum emission. }
\begin{tabular}{ c|c|c|c } 
  \hline
  \hline
\noalign{\smallskip}
           & Property      & Best value $\pm1\sigma$ & unit \\
\noalign{\smallskip}
  \hline
\noalign{\smallskip}
RW\,Aur\,A &$\Delta$RA\,(SB1) & $ 0.65 \pm 0.02 $ & mas \\
position   &$\Delta$Dec\,(SB1)& $-0.14 \pm 0.01 $ & mas \\
relative to&$\Delta$RA\,(SB2) & $-133.50 \pm 10.00 $  & mas \\
phasecenter&$\Delta$Dec\,(SB2)& $ 6.11 \pm 12.00 $  & mas \\
           &$\Delta$RA\,(SB3) & $ 0.44 \pm 0.04 $ & mas \\
           &$\Delta$Dec\,(SB3)& $ 0.18 \pm 0.12 $ & mas \\
           &$\Delta$RA\,(SB4) & $ 1.41 \pm 0.09 $ & mas \\
           &$\Delta$Dec\,(SB4)& $ 0.41 \pm 0.07 $ & mas \\
           &$\Delta$RA\,(LB1) & $ 0.12 \pm 0.05 $ & mas \\
           &$\Delta$Dec\,(LB1)& $ 3.16 \pm 0.06 $ & mas \\
\noalign{\smallskip}
  \hline
\noalign{\smallskip}
Flux       & $i_{SB1}$  & $ 1.063 \pm 0.001 $ & - \\
Amplitude  & $i_{SB2}$  & $ 1.055 \pm 0.006 $ & - \\
Scaling    & $i_{SB3}$  & $ 1.006 \pm 0.002 $ & - \\
           & $i_{LB1}$  & $ 0.970 \pm 0.001 $ & - \\
\noalign{\smallskip}
  \hline
\noalign{\smallskip}
RW\,Aur\,A & $f_{A0}$     & $ 522.08_{-28.29}^{+25.39} $ & $\mu$Jy/pix \\
profile    & $f_{A1}$     & $  14.19 \pm 0.41 $ & $\mu$Jy/pix \\
           & $\sigma_{A1}$& $  14.48 \pm 0.38 $ & mas \\
           & $f_{A2}$     & $   1.27 \pm 0.02 $ & $\mu$Jy/pix \\
           & $r_{A2}$     & $ 129.46 \pm 0.25 $ & mas \\
           & $\alpha_A$   & $  -0.44 \pm 0.01 $ & - \\
           & $\beta_A$    & $ -22.07 \pm 0.62 $ & - \\
\noalign{\smallskip}
  \hline
\noalign{\smallskip}
RW\,Aur\,B & $f_{B1}$     & $ 9.91_{-1.80}^{+1.38} $ & $\mu$Jy/pix \\
profile    & $r_{B1}$     & $ 41.49 \pm 0.40 $ & mas \\
           & $\sigma_{B1}$& $  0.68 \pm 0.31 $ & mas \\
           & $f_{B2}$     & $ 1.73_{-0.03}^{+0.03} $ & $\mu$Jy/pix \\
           & $r_{B2}$     & $ 40.07 \pm 1.32 $ & mas \\
           & $\sigma_{B2}$& $ 30.81 \pm 0.93 $ & mas \\
\noalign{\smallskip}
  \hline
  \hline
\end{tabular}
\tablefoot{ \centering The pixel size in the model image was 4\,mas. Observation SB2 is from the ACA array. }
\label{tab:uv_mcmc}
\end{table}

\section{ Orbital solutions \label{app:sec:hyperbolic} }

The package \texttt{rebound} \citep{rein2012, rein2015} was used to calculate the coordinates of RW\,Aur\,B given a set of orbital parameters. This was done by including a point mass at the origin of the coordinate system, containing the combined mass of the binaries and massless particles representing the position of RW\,Aur\,B at the time of each observation. We did not utilize the simulation capabilities of \texttt{rebound}, which can evolve the positions running a numerical simulation. These massless particles shared the same orbital parameters and would only differ in their true anomaly, which we will refer to in this Section as ``$\nu$''. The value of $\nu$ represents the angular distance between the current particle position and the vector pointing from the origin to the periastron of the orbit. 

For a given set of orbital parameters, the value of $\nu$ can be measured for each astrometric measurement. However, describing it as a function of physical time $\nu:=\nu(t)$ is relevant to compare the astrometry of each observed epoch. Even though there are Python packages to calculate orbital positions as a function of time for elliptical orbits \citep[e.g., \texttt{orbitize!}, ][]{orbitize2020}, most of the currently available tools to calculate hyperbolic orbits are made to calculate orbits with $\nu$ as input. In the following, we summarize the steps we took to go from time to $\nu$. 

First, we calculate the mean angular motion ``$\mathsf{n}$'' as a function of the total mass of the system  ``$m_{\text{tot}}$'' and the hyperbolic semi-transverse axis ``$a$'': 
\begin{equation}
    \mathsf{n} \, = \, \sqrt{\frac{Gm}{-a^3}} \text{,}
    \label{app:eq:mean_angular_motion}
\end{equation}

\noindent where $G$ is the gravitational constant, and $a<0$ by definition. In our MCMC, the value of $a$ is calculated from $d_{peri}$ and $\text{ecc}$ as $a=d_{peri}/(1-\text{ecc})$. From $\mathsf{n}$, we can calculate the mean anomaly ``$\mathsf{M}$'', as:
\begin{equation}
    \mathsf{M} \, = \, \mathsf{n} \cdot (t - t_{\text{per}}) \text{,}
    \label{app:eq:mean_anomaly}
\end{equation}

\noindent where $t_{\text{per}}$ is the physical time of periastron, and $t$ is the time of the measurement. The units of time of $t$ must be the same as those of $G$. Next, the hyperbolic anomaly ``$\mathsf{F}$'' of a measurement is related to $\mathsf{M}$ by following:
\begin{equation}
    \mathsf{M} \, = \, (\text{ecc} \cdot \sinh{(\mathsf{F})}) \cdot \mathsf{F} \text{,}
    \label{app:eq:hyperbolic_anomaly}
\end{equation}

\noindent which has no analytical solution when solving $\mathsf{F}$ from $\mathsf{M}$. Instead, we solve this equation numerically by using the Newton's method, where:
\begin{equation}
    \mathsf{F}_{k+1} \, = \, \mathsf{F}_k \, + \, \frac{\mathsf{M} \, - \, \text{ecc} \cdot \sinh{(\mathsf{F}_k)} \, + \, \mathsf{F}_k} { \text{ecc} \cdot \cosh{(\mathsf{F}_k)} \, - \, 1 } \text{,}
    \label{app:eq:newtons_method}
\end{equation}

\noindent which we iterate in $k$ until the difference $|\mathsf{F}_{k+1}-\mathsf{F}_k|$ is below $10^{-8}$\,deg. We use as initial guess $\mathsf{F}_0=\mathsf{M}$. From the hyperbolic anomaly $\mathsf{F}$, we finally derive the true anomaly $\nu$ as: 
\begin{equation}
    \nu \, = \, 2 \cdot \text{arctan}\left( \sqrt{\frac{\text{ecc} + 1}{\text{ecc} -1}} \cdot \text{tanh}\left(\frac{\mathsf{F}}{2}\right) \right) \text{,}
    \label{app:eq:true_anomaly}
\end{equation}

\noindent thus completing the transformation of time at observation of each astrometric measurement to true anomaly $\nu$, which is used as input to recover the coordinates and velocity at each point of the hyperbolic orbit. 

For consistency, we also used \texttt{rebound} to calculate the elliptical orbits. Thus, we follow a similar procedure to go from $\nu$ to time. The mean angular motion is now calculated as: 
\begin{equation}
    \mathsf{n} \, = \, \sqrt{\frac{Gm}{a^3}} \text{,}
    \label{app:eq:mean_angular_motion2}
\end{equation}

\noindent which is used to obtain the mean anomaly $\mathsf{M}$. The relation between the elliptical anomaly $\mathsf{E}$ and $\mathsf{M}$ is: 
\begin{equation}
    \mathsf{M} \, = \, \mathsf{E} - (\text{ecc} \cdot \sin{(\mathsf{E})}) \text{,}
    \label{app:eq:elliptical_anomaly}
\end{equation}

\noindent which we also solve with the Newton's method. This $\mathsf{E}$ is used to obtain the true anomaly $\nu$. 

It is relevant to note that the semi-major axis for elliptical orbits, and semi-transverse axis for hyperbolic orbits, is a quantity that diverges for $\text{ecc} = 1$. Thus, we chose to run the MCMC with the distance at periastron instead of the semi-major (semi-transverse) axis, which are related through the eccentricity by: 
\begin{equation}
    d_{\text{per}} \, = \, a \cdot (1 - \text{ecc}) \text{.}
    \label{app:eq:dist_periastron}
\end{equation}

\section{Comparison of elliptical and hyperbolic orbits}

\begin{figure}
    \centering
    \includegraphics[width=9cm]{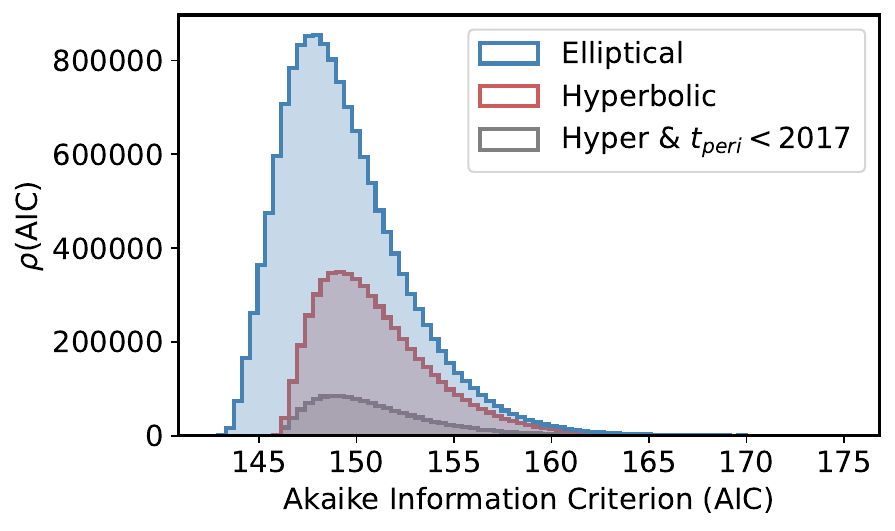}\\
    \caption{ Histogram of the AIC value for all walkers, separated by elliptical, hyperbolic, and hyperbolic orbits where the periastron time is before the first ALMA observation. } 
    \label{app:fig:statistical_tests}
\end{figure}

In Fig.~\ref{app:fig:statistical_tests}, the histograms of the AIC (Akaike Information Criterion) is shown for the walkers with $\text{ecc}<1$ (elliptical) and $\text{ecc}>1$ (hyperbolic). 

\section{Predicted astrometry}

\begin{figure}
    \centering
    \includegraphics[width=9cm]{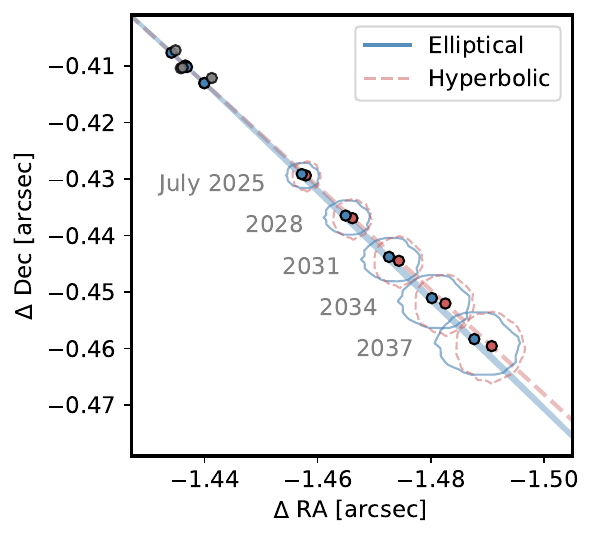}\\
    \caption{ Predicted location of RW\,Aur\,B based on the elliptical (solid blue) and hyperbolic (dashed red) models, starting from July 2025, and then the July every three years until 2037. The position from the best elliptical and hyperbolic model are shown with a circles, while the contours show the $3\sigma$ scatter for each family or orbits. } 
    \label{app:fig:future}
\end{figure}

The orbital solutions derived in Sect.~\ref{sec:rwaurb_orbit} allow us to predict the future position of RW\,Aur\,B, which we show in Fig.~\ref{app:fig:future}. Given the current astrometric measurements, a single astrometric measurement will most likely not be enough to distinguish between elliptical and hyperbolic solutions. Several measurements over the years will be needed to constrain the nature of the RW\,Aur orbit. 

\section{RW\,Aur CO isotopologues emission}

\begin{figure*}[t]
 \centering
        \includegraphics[width=18cm]{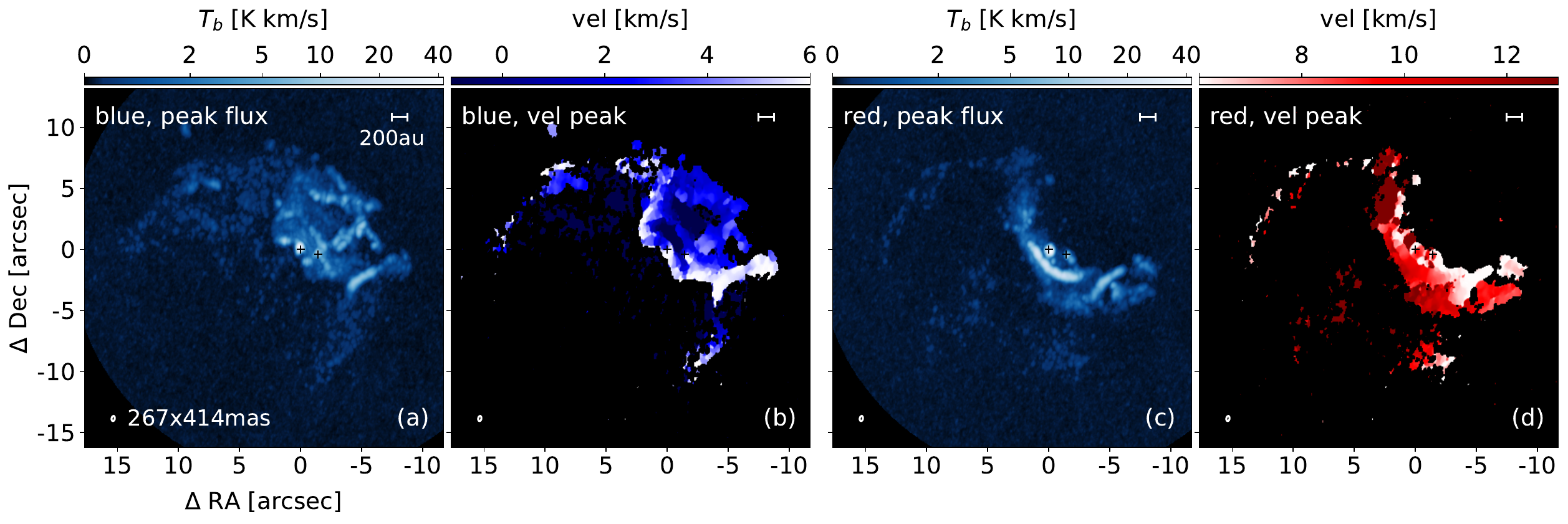}\\
        \vspace{-0.1cm}
   \caption{The $^{12}$CO blueshifted emission with respect to RW\,Aur\,A is shown in panels (a) and (b), while the redshifted component is shown in (c) and (d). The beam size in the lower left corner and the scale bar in the upper right are the same for all the panels. The centers of the 2 disks are shown with black $+$ symbols.}
   \label{fig:bluered}
\end{figure*}

\begin{figure*}[t]
 \centering
        \includegraphics[width=18cm]{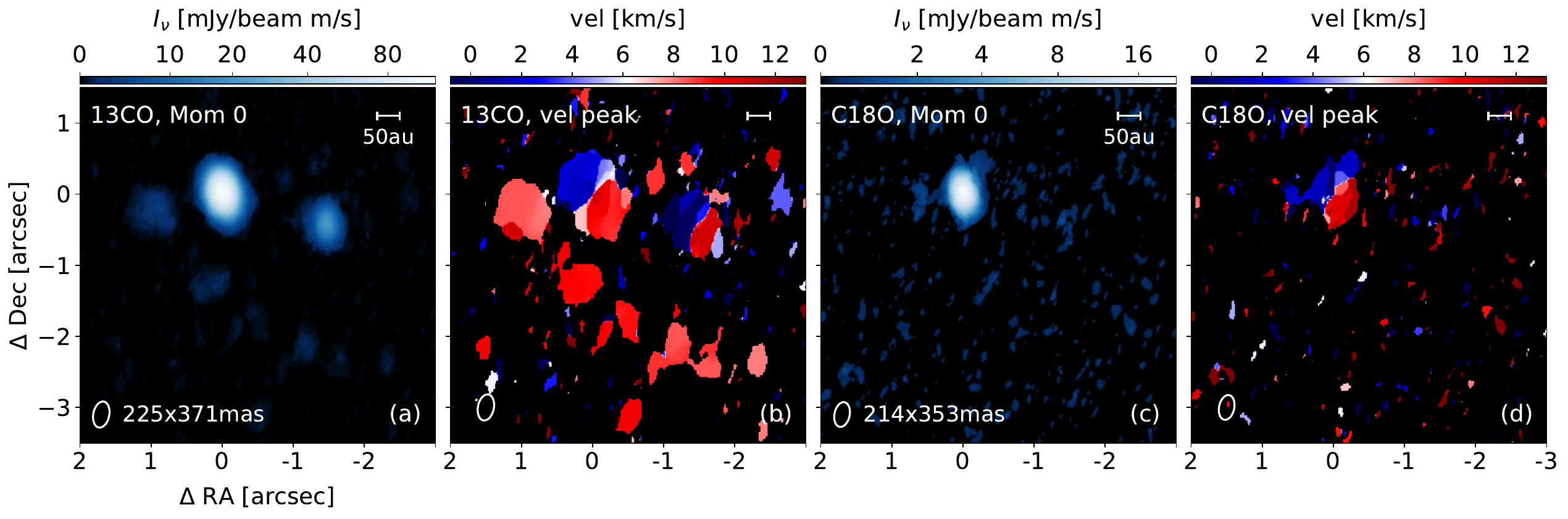}\\
        \vspace{-0.1cm}
   \caption{$^{13}$CO Moment 0 and velocity at peak flux are shown in panels (a) and (b), while the same images for C$^{18}$O are shown in panels (c) and (d). Beam sizes are shown in lower left corner, and a scale bar of 20\,au is shown in the upper right corner.}
   \label{fig:co_iso}
\end{figure*}

The velocity map of RW\,Aur CO isotopologues are shown in Fig.~\ref{fig:bluered} and Fig.~\ref{fig:co_iso}. The $^{13}$CO J=2-1 emission is detected in both disks, and the bright tidal arm to the south of RW\,Aur\,A is detected too. In C$^{18}$O J=2-1 emission, only RW\,Aur\,A is detected. Due to the moderate angular resolution of these detections, it is not possible to explore the radial morphology of the emission with the same detail as the dust continuum or $^{12}$CO J=2-1 emission.

\end{appendix}

\end{document}